\documentstyle[epsf,11pt]{article}
\newcommand{\be}{\begin{eqnarray}}

\newcommand{\ee}{\end{eqnarray}}
\newcommand{\munu}{\mu\nu}
\newcommand{\psibar}{\bar \psi}

\title{
        \begin{flushright}
        {\normalsize NBI--98--25\\
        TPI--MINN--98/16--T \\
        NUC--MINN--98/8--T \\
        UMN-TH-1716-98 \\
        September 1998 \\}
        \end{flushright}
\bf  Fock Space Distributions, Structure Functions, Higher Twists and
Small x}
\author{
Larry McLerran\\
       {\small\it Theoretical Physics Institute, University of
Minnesota,
        Minneapolis, MN 55455} \\
        Raju Venugopalan \\
        {\small\it Niels Bohr Institute,
        Blegdamsvej 17,
        Copenhagen, Denmark, DK--2100 } \\
         }
\date{}

\begin{document}

\maketitle

\begin{center}
{\bf Abstract}\\
\end{center}

We compute quark structure functions and the intrinsic Fock space
distribution of sea quarks in a hadron wavefunction at small x. The
computation is performed in an effective theory at small x where the
gluon field is treated classically. At $Q^2$ large compared to an
intrinsic scale associated with the density of gluons $\mu^2$, large
compared to the QCD scale $\Lambda^2_{QCD}$, and large compared to the
quark mass squared $M^2$, the Fock space distribution of quarks is identical
to the distribution function measured in deep inelastic scattering.
For $Q^2 \le M^2$ but $Q^2 >> \mu^2$, the quark distribution is
computed in terms of the gluon distribution function and explicit
expressions are obtained.  For $Q^2 \le \mu^2$ but $Q^2 >>
\Lambda_{QCD}^2$ we obtain formal expressions for the quark
distribution functions in terms of the glue. An evaluation of these
requires a renormalization group analysis of the gluon distribution
function in the regime of high parton density.  For light quarks at
high $Q^2$, the DGLAP flavor singlet evolution equations for the
parton distributions are recovered.  Explicit expressions are given
for heavy quark structure functions at small x.

\vfill \eject

\section{Introduction}

One of more interesting problems in perturbative QCD is the behaviour
of structure functions at small values of Bjorken x. In deep inelastic
scattering (DIS) for instance, for a fixed $Q^2>>\Lambda_{QCD}^2$, the
operator product expansion (OPE) eventually breaks down at sufficiently small
x~\cite{AMueller1}. Therefore at asymptotic energies, the conventional
approaches towards computing observables based on the linear
DGLAP~\cite{DGLAP} equations are no longer applicable and novel
techniques are required. Even at current colider energies such as 
those of HERA where the conventional wisdom is that the DGLAP equations
successfully describe the data, there is reason to believe that
effects due to large logarithms in $\alpha_S\log(1/x)$ (or large
parton densities) are not small and we may be at the threshold of a
novel region where non--linear corrections to the evolution equations
are large~\cite{Levin,FrankStrik1}. One reason violations of DGLAP evolution 
have not been seen clearly thus far at HERA is the small phase volume for
gluon emission~\cite{FrankStrik2}. A straightforward way to further probe this
region at current colider energies would be by using nuclear beams at
HERA~\cite{Strikman} or in electron--proton collisions at the LHC 
collider where the phase volume for gluon emission is signficantly 
larger~\cite{FELIX}.

In recent years, a non--OPE based effective field theory approach to
small x physics has been developed by Lipatov and
collaborators~\cite{Lipatov}. Their initial efforts resulted in an
equation known popularly as the BFKL equation~\cite{BFKL}, which sums
the leading logarithms of $\alpha_S\log(1/x)$ in QCD. In marked
contrast to the leading twist Altarelli--Parisi equations for
instance, it sums all twist operators that contain the leading
logarithms in x. The solutions to the BFKL equation predict a rapidly
rising gluon density and there was much initial euphoria when the H1
and ZEUS data at HERA showed rapidly rising parton
densities~\cite{H1ZEUS}. However, it was shown since:
\begin{itemize}
 \item
The rapid rise of the structure functions can arguably be accounted for
by
the next to leading order (NLO) 
DGLAP equations by appropriate choices of initial
parton densities~\cite{GRV, CTEQ}.

\item
 The next to leading
logarithmic corrections to the BFKL equation computed in the above
mentioned effective field theory approach are
{\it very} large~\cite{FadLip,CamCia,KovMuell}.
\end{itemize}
As a consequence, the theoretical
situation is wide open and novel approaches need to be explored.

An alternative effective field theory approach to QCD at small x was
put forward in a series of papers~\cite{MV}-\cite{JKW}. In the
approach of Lipatov and collaborators, the fields of the effective theory
are composite reggeons and pomerons. This approach is motivated by the 
reggeization of the gluon that occurs in the leading log result. Our
approach based on Refs.\cite{MV}-\cite{JKW} is instead a 
Wilson renormalization group approach where the fields are
those of the fundamental theory but the form of the action at small x is
obtained by integrating out modes at higher values of x. Integrating out the
higher x modes results in a set of non--linear renormalization group
equations~\cite{JKW}. If the parton densities are not too high, the 
renormalization group equations can be linearized and have been shown to 
agree identically with
the leading log BFKL and small x DGLAP equations~\cite{JKLW}. There is
much effort underway to explore and make quantitative predictions for the
non--linear regime beyond~\cite{JKW2,ILM}.

In this paper, we apply the above effective action approach to study
the fermionic degrees of freedom at small x. At small values of x, gluon
degrees of freedom dominate and the fermionic degrees of freedom
present are essentially the sea quarks that are radiatively generated
from the glue~\cite{Al} (and are therefore $O(\alpha_S)$ corrections).
Nevertheless, the sea quark distributions are extremely important
since they are directly measured in deep inelastic scattering
experiments. In this paper, we will develop a formalism, in the context
of the renormalization group approach, which relates
structure functions at small x to the sea quark distributions, and
therefore to the gluon distribution. We derive analytic expressions
summing a particular class of all twist operators which we argue give
the
dominant contribution at small x. At leading twist, in light cone gauge,
these reduce to the well known simple relation between the structure
function
$F_2$ and the sea quark Fock space distribution function~\cite{Jaffe}
\be
F_2(x,Q^2) = \int_0^{Q^2} dk_t^2 {dN_{sea}\over {dk_t^2 dx}}
\, .\nonumber
\label{leading twist}
\ee
Above, x and $Q^2$ are the usual invariants in deep inelastic
scattering. We show explicitly that for light quarks and at high
$Q^2$, we reproduce the Altarelli--Parisi evolution equations for the
quark distributions at small x.

There has been much interest in heavy quark distributions motivated
partly by the significant contribution of heavy quarks to the structure
functions at HERA~\cite{HERAcharm}. Until recently, heavy quarks were treated 
as infinitely massive for $Q^2$ equal to or less then the quark mass
squared and massless below. There now exist approaches which study quark
distributions in a unified manner for a range of $Q^2$ and quark masses
(for a discussion and further references see Refs.~\cite{ACOT,Thorne}).
A nice feature of our formalism is that heavy quark evolution is treated
on the same footing as light quarks and specific predictions can be made for
$F_2^{charm/bottom}/F_2$ within our formalism. These can also be related
to the diffractive cross section at small x but that issue will not be
addressed in this paper. This issue and the relation of our work to 
the above cited work and other recent works on heavy quark production at small
x~\cite{Smith,MartinRobSterl} will be addressed at a later date.

The results of our analysis are the following. In our theory of the
gluon distribution functions, a dimensionful scale appears which
measures the density of gluons per unit area,
\be
   \mu^2 = {1 \over \sigma} {{dN} \over {dy}} \, ,
\ee
where $\sigma$ is the hadronic cross section of interest.   Here $y =
y_0 - ln(1/x)$, $y_0$ is an arbitrarily chosen constant and $x$ is
Bjorken $x$.  When this parameter satisfies $\mu >> \Lambda_{QCD}$
the gluon dynamics, while nonperturbative, is both weakly coupled and
semiclassical.  We shall always assume that at sufficiently small x
this is satisfied.

At small x, the gluon field, being bosonic, has to be treated
non-perturbatively.  This is analogous to the strong field limit used in
Coulomb problems.  Fermions, on the other hand, do not develop a large
expectation value and may be treated perturbatively.  To lowest order in
$\alpha_{QCD}$, the gluon distribution function is determined by
knowing the fermionic propagator in the classical gluon background
field. 
In general, this propagator must be determined to all orders in the
classical gluon field as the field is strong.  This can be done due to
the simple structure of the background field. 

At this point it is useful to distinguish between two 
different quantities which are often used interchangeably.  
One is the quark structure functions as measured
in deep inelastic scattering and the second is the Fock space
distribution of quarks.  At high $Q^2$ which is usually the case
considered in perturbative QCD, these two quantities are essentially
identical.  However, for massive quarks when $Q^2 \le M^2$ even
if $Q^2 >> \Lambda_{QCD}^2$, the two quantities differ.  
For $Q^2 >> \mu^2$, regardless of mass,
the quark Fock space distribution function and the quark structure
functions may both be simply expressed as linear functions of the gluon
distribution functions.  Therefore, with knowledge of the gluon
distribution function, one can compute the quark distribution function.

For the case of massless quarks,  when $\mu^2 \ge Q^2 >>
\Lambda_{QCD}^2$, we are still in the weak coupling limit.  However, we
must keep all orders in the gluon field.  In this region, the
integration over the gluon fields in our effective field theory
cannot be directly performed as yet since it requires a full renormalization
group analysis of the theory. In other words,
the measure of integration for the high density regime
in the effective theory 
has not yet been computed.  Nevertheless, we obtain an explicit
functional dependence on the gluon fields which must be integrated over 
with the right measure.

The power of the technique which we use to analyze this problem  
is that it does not rely on a high twist expansion.  It uses only the weak
coupling nature of the theory which must be true at small x if the gluon
density is very high.  We are therefore in a position to find
non-trivial relations between these various parameters in a region where 
the weak coupling analysis is valid but where perturbation theory and
leading order operator product expansion methods are not valid.

This paper is organized as follows. In section 2 we write down and
review an effective action for the small x modes in QCD. The action is
imaginary and the modes are averaged over with a statistical weight
\be
\exp\left[-F[\rho]\right]\nonumber \, ,
\ee
where $F[\rho]$ is a functional over the color charge density $\rho$
of the higher x modes. The functional $F[\rho]$ obeys a non--linear
renormalization group equation. Of particular interest in this paper
is the saddle point solution of the effective action since the sea
quark distributions are computed in the classical background field of
this action.    This section is a quick review of known results which are
necessary to understand the remainder of the paper.

An expression relating the electromagnetic
current--current correlator to the fermion propagator in the classical
background field is derived in section 3. We also discuss the light
cone Fock space distributions and their relation to the structure
functions in this section.

In section 4, we solve the Dirac equation in
the classical background field and obtain an explicit expression for
the fermion Green's function in the classical background field.

The Green's function is used in section 5 to compute the sea quark
distribution function and the leading twist contribution to the structure
function $F_2 (x,Q^2)$ at small x. The color averaging over the functional 
$F[\rho]$ in the distribution function is compactly represented by a 
function ${\tilde \gamma}(p_t)$, where $p_t$ can be interpreted as the 
intrinsic transverse momentum of the glue at high parton densities.
It is shown explicitly that for
light quark masses the flavor singlet Altarelli--Parisi equations at
small x are recovered.

The current--current correlator and the
structure functions are computed explicity in section 6. As a check of
our computation, it is shown that the leading twist results are
recovered in the appropriate limit. The heavy quark structure
functions are computed explicitly. The phenomenological implications and
the connections to the recent literature on heavy quark production will
not be addressed in this work but will be considered at a later date.

Section 7 contains a summary of our results and a discussion of future
work.

The first of two appendices contains a discussion of our notation and
conventions. In the second appendix we present an explicit form for
${\tilde\gamma} (p_t)$ for the particular case of Gaussian color
fluctuations.

\section{Effective Field Theory for Small x Partons in QCD: Review of
Results}
\vskip 0.15in

We will discuss below an effective action for the wee parton modes in
QCD.
The action contains an imaginary piece which involves functional
$F[\rho]$ which satisfies a
non-linear Wilson renormalization group equation. In the weak field
limit of
this renormalization group equation, the BFKL equation is recovered.
In the double logarithmic region, the evolution equation is also
equivalent to DGLAP~\footnote{This was first noticed by Yuri Dokshitzer in
his paper in Ref.~\cite{DGLAP}.}.  We
next discuss the classical background field which is the saddle point
solution of this action. It is this background field that the sea quarks
couple to at small x and the properties of the background field will be
relevant for the discussion in later sections.

\subsection{The Effective Action and the Wilson Renormalization Group at
Small x}
\vskip 0.15in

In the infinite momentum frame $P^+\rightarrow \infty$, the effective
action
for the soft modes of the gluon field with longitudinal momenta
$k^+<<P^+$ (or equivalently
$x\equiv k^+/P^+ << 1$) can be written in light cone gauge $A^+=0$ as
\be
S_{eff} &=& -\int d^4 x {1\over 4} G_{\munu}^{a}G^{\munu,a} +{i\over N_c}
\int d^2 x_t dx^- \rho^a(x_t,x^-)
{\rm Tr}\left(\tau^a W_{-\infty,\infty}[A^-](x^-,x_t)\right)\nonumber \\
&+& i\int d^2 x_t dx^- F[\rho^a(x_t,x^-)] \, .
\label{action}
\ee
Above, $G_{\munu}^a$ is the gluon field strength tensor, $\tau^a$ are
the
$SU(N_c)$ matrices in the adjoint representation and $W$ is the path
ordered
exponential in the $x^+$ direction in the adjoint representation of
$SU(N_c)$,
\be
W_{-\infty,\infty}[A^-](x^-,x_t) = P\exp\left[-ig\int dx^+
A_a^-(x^-,x_t)\tau^a\right] \, .
\ee
The above is the most general gauge invariant form~\cite{JKLW} of the
action
that was proposed in Ref.~\cite{MV}.  

This is an effective action valid in a limited range of  $P^+ <<
\Lambda^+$ where $\Lambda^+$ is an ultraviolet cutoff in the plus
component of the momentum.  The degrees of freedom at higher values of
$P^+$ have been integrated out and their effect is to generate the
second and third terms in the action.

The first term in the above is the usual
field strength piece of the QCD action and describes
the dynamics of the wee partons at the small x values of interest. The
second term
in the above is the coupling of the wee partons to the hard color
charges at higher rapidities, with x values corresponding to values of
$P^+ \ge \Lambda^+$.   When expanded to first order in $A^-$ this term
gives the ordinary $J \cdot A$ coupling for classical fields.  The
higher order terms are needed to ensure a gauge invariant coupling of
the fields to current. 

In the infinite
momentum frame, only the $J^+$ component of the current is large (the
other components being suppressed by $1/P^+$). The longer wavelength wee
partons do not resolve the higher rapidity parton sources to within
$1/P^+$
and for all practical purposes, one may write
\be
\rho^a (x_t,x^-)\longrightarrow \rho^a (x_t) \delta(x^-) \, .
\ee
The last term in the effective action is imaginary. It can be thought of
as a statistical weight resulting from integrating out the higher
rapidity modes in the original QCD action. Expectation values of gluonic
operators $O(A)$ are then defined as
\be
<O(A)> = { \int [d\rho] \exp\left(-F[\rho]\right) \int [dA] O(A)
\exp\left(iS[\rho,A]\right) \over{\int [d\rho] \exp\left(-F[\rho]\right)
\int [dA] \exp\left(iS[\rho,A]\right)}} \, ,
\label{expvalue}
\ee
where $S[\rho,A]$ corresponds to the first two terms in
Eq.~\ref{action}.
The color averaging procedure for fermionic observables is discussed
further in section 5.1.

In Ref.~\cite{MV} a Gaussian form for the action
\be
\int d^2 x_t {1\over 2\mu^2} \rho^a \rho^a \, ,
\label{Gauss}
\ee
was proposed, where $\mu^2$ was the average color charge squared per
unit
area of the sources at higher rapidities than is appropriate for our
effective action, that is. For large nuclei $A>>1$ it was shown
that
\be
      \mu^2= {1 \over { \pi R^2}} {N_q \over {2N_c}} \sim A^{1/3}/6
\,\mbox{fm}^{-2}.
\ee 
This result was independently confirmed in a
model constructed in the nuclear rest frame~\cite{Kovchegov}.
If we include the contribution of gluons which have been integrated out by the
renormalization group technique, one finds that~\cite{gyulassy}
\be
        \mu^2 = {1 \over {\pi R^2}} \left( {N_q \over {2N_c} }+ {{N_cN_g} \over
{N_c^2-1}} \right)
\ee
Here $N_q$ is the total number of quarks with x above the cutoff
\be
        N_q = \sum_i \int_x^1 dx^\prime q_i(x^\prime)
\ee
where the sum is over different flavors, spins, quarks and antiquarks. 
For gluons, we also have
\be
        N_g = \int_x^1 dx^\prime g(x^\prime)
\ee
The value of $\pi R^2$ is well defined for a large nucleus.  For a
smaller hadron, we must take it to be $\sigma$, the total cross section
for hadronic interactions at an energy corresponding to the cutoff. 
This quantity will become better defined for a hadron in the
renormalization group analysis.   

The above equation for $\mu^2$ is subtle because, implicitly, on the right
hand side, there is a dependence on $\mu$ through the structure functions
themselves.  This is the scale at which they must be evaluated. 
Calculating $\mu$ therefore involves solving an implicit equation.
Note that because the gluon distribution function rises rapidly
at small x, the value of $\mu$ grows as x decreases. 

The Gaussian
form of the functional $F[\rho]$ is reasonable when the color charges at
higher rapidity are uncorrelated and are random sources of color charge.
This is true for instance in a very large nucleus.  It is also true if
we study the Fock space distribution functions or deep inelastic
structure functions at a transverse momentum scale which is larger than
an intrinsic scale set by $\alpha_S \mu$.  In this equation
$\alpha_S$ is evaluated at the scale $\mu$.  At smaller transverse
momenta scales, one must do a complete renormalization group analysis to
determine $F[\rho]$.  This analysis is not yet complete, but should be
feasible in the context of the weakly coupled field theory as long as
the transverse momentum scale remains larger than that $\Lambda_{QCD}$.
The color averaging procedure for the case of Gaussian fluctuations is
discussed in appendix B.

A final comment about the limit of applicability of the classical action
above concerns limitations in the transverse momentum range.  The action
above is valid only when probing transverse momenta scales $p_T \le
\mu$.  This includes the Gaussian region since $\alpha_S << 1$.  At
higher transverse momenta, one must use DGLAP evolution, with the
structure functions as determined at lower values of $Q^2$ as boundary
conditions~\cite{RajLar}. 
This will be important for the case of heavy quarks as the 
transverse momentum scale there is very large.

The above comments on the renormalization group analysis show the
limitations of our analysis with respect to quarks.  For transverse
momentum scales $p_t >> \alpha_s \mu$, one can use a Gaussian source and
all relevant quantities can be computed explicitly. 
At smaller scales, one can derive a formal expression,
which, hopefully, will be directly computed in the near future.
For heavy quarks, the Gaussian analysis should be adequate.

To complete a review of the renormalization group, we briefly review the
procedure used to determine $F[\rho]$.   It was shown that a Wilson
renormalization
group procedure~\cite{JKMW} could be applied to derive a non-linear
renormalization group equation for $F[\rho]$. The procedure, briefly, is
as follows. The gauge field is split as
\be
A_\mu^a(x) = b_\mu^a(x) + \delta A_\mu^a(x) + a_\mu^a(x) \, ,
\ee
where $b_\mu^a(x)$ is the saddle point solution of Eq.~\ref{action} and
corresponds to the hard modes above the longitudinal momentum scale
$\Lambda$. The fluctuation field $\delta A_\mu^a(x)$ contains the soft
modes
$\Lambda^{\prime+} < k^+<\Lambda^+$ and $a_\mu(x)$ are soft fields with
longitudinal momenta $k^+<\Lambda^{\prime+}$. The cutoffs are chosen
such that
$\alpha_S \log(\Lambda^{\prime+}/\Lambda^+)<<1$. Small fluctuations are
performed
about the saddle point solution $b_\mu^a(x)$ to the effective action at
the scale $\Lambda^+$, to obtain the effective action for the fields
$a_\mu^a(x)$ at the scale $\Lambda^{\prime+}$. The new charge density at
this
scale $\rho^\prime$ is given by $\rho^{a \prime} = \rho^a + \delta
\rho^a$,
where $\delta \rho^a$ can be expressed as the sum of linear and
bi-linear
terms in the fluctuation field $\delta A_\mu^a(x)$.

To leading order in
$\alpha_S$, the effective action at the scale $\Lambda^{\prime+}$ can be
expressed in the same form as Eq.~\ref{action}, with the functional
$F[\rho^\prime]$ satisfying a non--linear renormalization group
equation~\cite{JKLW}-\cite{JKW}.
In terms of the statistical weight $Z=\exp(-F[\rho])$, it can be
expressed as
\be
{d Z\over {d \log(1/x)}}=\alpha_S \left[{1\over 2} {\delta^2\over
{\delta \rho_\mu \delta \rho_\nu}}\left(Z\chi_{\munu}\right)
-{\delta\over
{\delta\rho_\mu}} \left(Z\sigma_\mu\right)\right] \, ,
\label{nonlinRG}
\ee
where $\sigma[\rho]$ and $\chi[\rho]$ are respectively one and two point
functions obtained by integrating over $\delta A$ for fixed $\rho$. The
one point function $\sigma$ includes the virtual corrections to
$F[\rho]$
while the two point function $\chi$ includes the real contributions to
$F[\rho]$. Both of these can be computed explicitly from the small
fluctuations propagator in the classical background field. The
propagator
was first computed by fixing the residual gauge freedom
to be $\delta A^-(x^-=0)$~\cite{AJMV} but a less restrictive gauge
choice was later found which may be useful for computing 
$\sigma$ and $\chi$~\cite{JKW}.

For weak fields, the free gluon propagator can be used to obtain the
well known BFKL equation for the unintegrated gluon density, which is
defined as
\be
{dN\over {dk_t^2}} = \int d^2x_t e^{-ik_t x_t} <\rho(x_t)
\rho(0)>_\rho \, .
\ee
Performing the renormalization group procedure defined above to obtain
the charge density $\rho^\prime = \rho + \delta \rho$, one obtains
\be
<\rho^\prime \rho^\prime>_\rho - <\rho \rho>_\rho =
\alpha_S \log(1/x) \left[ 2<\rho \sigma>_\rho + <\chi>_\rho \right] \, .
\label{colorchange}
\ee
The one and two point functions $\sigma$ and $\chi$ respectively can
be computed to linear order in the classical background field and the
results are~\cite{JKLW}
\be
\sigma^a(k_t) = - {g^2 N_c\over {2(2\pi)^3}}\rho^a(k_t) \int
d^2 p_t {k_t^2\over {p_t^2 (p_t-k_t)^2}} \, ,
\ee
and
\be
\chi = {2g^2 N_c \over {(2\pi)^3}} \int d^2 p_t \rho^a(p_t) \rho^a(-p_t)
{k_t^2\over {p_t^2 (k_t-p_t)^2}} \, .
\ee
The above are respectively the virtual and real contributions to the
change
in the color change density after integrating out the modes
$\Lambda^{\prime+}
< k^+ < \Lambda^+$. Substituting these into Eq.~\ref{colorchange}, one
obtains the well known BFKL equation
\be
x{dN\over {dk_t^2 dx}} = {\alpha_S N_c\over {2\pi^2}} \int d^2 p_t
{k_t^2\over {p_t^2 (p_t-k_t)^2}}\left[{dN\over {dk_t^2}} - 2 {dN\over
{dp_t^2}}\right] \, .
\ee

Strenuous efforts are currently underway to compute $\sigma$ and $\chi$
to
all orders in the background field and thereby solve the full
non--linear
Wilson renormalization group equation for $F[\rho]$~\cite{JKW2,ILM}.

\subsection{The Classical Background Field at Small x}
\vskip 0.15in

The effective action in Eq.~\ref{action} has a remarkable saddle point
solution~\cite{MV,JKMW,Kovchegov}. It is equivalent to solving the Yang--Mills
equations
\be
D_\mu G^{\munu} = J^\nu \delta^{\nu +} \, ,
\ee
in the presence of the source $J^{+,a} = \rho^a(x_t,x^-)$.
Here we will allow the source to be smeared out in $x^-$ as this is
useful in the renormalization group analysis. It is also useful for intuitively
understanding the nature of the field.
One finds a solution where $A^{\pm}=0$ and
\be
A^i = {-1\over {ig}}\,V\partial^i V^\dagger  \, ,
\ee
for $i = 1,2$
is a pure gauge field which satisfies the equation
\be
D_i {d A^i\over dy} = g \rho (y,x_\perp) \, .
\ee 
Here $D_i$ is the covariant derivative $\partial_i + V\partial_i
V^\dagger$
and $y=y_0 + \log(x^-/x_0^-)$ is the space--time rapidity and $y_0$ is
the
space-time rapidity of the hard partons in the fragmentation region. At
small x we will use the space--time and momentum space notions of
rapidity interchangeably~\cite{RajLar}.  The momentum space rapidity is
defined to be $y = y_0 - ln(1/x)$ where $x$ is Bjorken $x$.
The solution of the above equation is
\be
A_\rho^i(x_t) = {1\over ig}\left(Pe^{ig\int_y^{y_0}
dy^\prime
{1\over {\nabla_{\perp}^2}}\rho(y^\prime,x_t)}\right)
\nabla^i\left(Pe^{ig\int_y^{y_0} dy^\prime {1\over
{\nabla_{\perp}^2}}\rho(y^\prime,x_t)}\right)^\dagger \, .
\label{puresoln}
\ee

To compute the classical nuclear gluon distribution function, for
instance, 
\be
{dN\over {d^3 k}} = {1\over {(2\pi)^3}} 2 |k^+| \int d^3 x d^3 x^\prime
e^{ik\cdot (x-x^\prime)} <A_i^a(x^-,x_t)A_i^a({x^\prime}^-,x_t^
\prime)>_{\rho} \,\, ,
\label{Green}
\ee
one needs in general to average over the product of the classical
fields at two space--time points with the weight $F[\rho]$ as shown in
Eq.~\ref{expvalue} or for the Gaussian measure with the weight in
Eq.~\ref{Gauss}. In the latter case, exact analytical solutions are
available for correlators in the classical background field. For the
case of interest here, the fermion Green's function in the classical
background field will depend on correlators of the form
$<V(x_t)V^\dagger(y_t)>_\rho$ 
where the $V$'s are $SU(N_c)$ gauge transformation matrices 
defined above.  In appendix B,
we discuss in detail the computation of this correlator for the case of
Gaussian fluctuations.

\section{The Current--Current Correlator at Small x}
\vskip 0.15in

In this section, we will derive a formal expression for the hadron
tensor
$W_{\munu}$ at small x relating it to the fermion Green's function in the
classical background field. We also derive a relation between the
light cone quark distribution function and the fermion Green's function.
To leading twist, the structure functions are simply related to the
light cone
quark distribution function. In general (for example, for heavy quark
distributions) this is not true. Nevertheless, a 
simple relationship may be 
found between the gluon distribution functions and that of the quarks.  
This is because
the quarks distribution functions is given by an integral over a propagator
in the classical gluon background field described above.

\subsection{Derivation}
\vskip 0.15in

In deep inelastic electroproduction, the hadron tensor can be expressed
in terms of the forward Compton scattering amplitude $T_{\munu}$ by
the relation~\cite{Pokorski}
\be
W^{\munu}(q^2,P\cdot q) &=& 2 Disc~ T^{\munu}(q^2,P\cdot q) \equiv
{1\over {2\pi}} {\rm Im} \int d^4 x \exp(iq\cdot x) \nonumber \\
&\times&  <P|T(J^\mu (x) J^\nu (0))|P> \, ,
\label{hadtensor}
\ee
where ``T'' denotes time ordered product,
$J^\mu={\psibar}\gamma^\mu \psi$ is the hadron
electromagnetic current and ``Disc'' denotes
the discontinuity of $T_{\munu}$ along its branch cuts in the variable
$P\cdot q$.  Also,
$q^2\rightarrow \infty$ is the momentum transfer squared of the virtual photon~\footnote{
Note that in our metric convention, a space--like photon has 
$q^2 = Q^2 > 0$.} and $P$ is the momentum of the target. 
In the infinite momentum frame,
$P^+\rightarrow \infty$ is the only large component of the momentum.
The fermion state above is used in the expectation value for the current operators
is normalized as 
$<P \mid P^\prime> = (2\pi)^3 E/m \,\delta^{(3)} (P-P^\prime)$
where $m$ is the mass of the target hadron.  This definition of $W^{\mu \nu}$
and normalization of the state is traditional, and 
we will abide by these conventions in spite
of the awkward factors of $m$.
We will see in the end that all factors of $m$ cancel from the definition 
of quantities of physical interest.  
(The normalization we will use in this paper
for quark and lepton states will have E/m replaced by $2P^+$.)

Let us first describe the computation of $<P\mid T(J^\mu(x) J^\nu(0))
\mid P>$.  In our computation, we have an external source corresponding
to the particle whose state vector is denoted by $\mid P>$.  Our source
is located at some fixed position.  We must therefore consider the
generalization of $W^{\mu \nu}$ for such a source which has a position
dependence.  Note that for a given source, we also have a lack of
translational invariance in the transverse direction.  Transverse
translational invariance is restored after integration over the source. 
There is no dependence of our source on $x^+$.  Therefore the relevant
variable is $x^-$.  

We now argue that the relevant definition of $W^{\mu \nu}$ is
\be
        W^{\mu \nu} (q^2, P\cdot q) = {1\over {2\pi}}\sigma {P^+ \over m} 
{\rm Im}  \int d^4x dX^- e^{i q\cdot x}
 < T\left(J^\mu(X^- +x/2) J^\nu (X^- -x/2)\right)>\, .\nonumber\\
\ee
To see this let us first verify that we can write $W^{\mu \nu}$ in this form
for the conventional definition valid for plane wave states $|P>$.  Notice
that we can define $< O > = <P \mid O \mid P> / <P \mid P>$ where $O$ is any
operator.  As mentioned above, the 
expectation value $<P \mid P> = (2\pi)^3 E/m\, \delta^{(3)} (0)
= (2\pi)^3 E/m ~V$. Here we shall take the spatial volume $V$ to
be $\sigma$ times an integral over the longitudinal extent of the state.
Using these conventions, we see that we reproduce the above definition of
$W^{\mu \nu}$.

This definition corresponds to treating the variable $X^-$ as a center
of mass coordinate and $x^-$ as a relative longitudinal position. For
a translationally invariant state, this would give the longitudinal dimension 
of the system.  
The definition is 
Lorentz covariant as will be shown explicitly in section 6. The
integration over $X^-$ is required since we must include all of the
contributions from quarks at all $X^-$ to the distribution function.
In our external source language, the variable $P^+$ can be taken to be the
longitudinal momentum corresponding to the fragmentation region.  In the
end all of the $P^+$ (and m) dependence will disappear upon taking the 
infinite momentum
limit.  To check this definition later, we shall show that it reproduces
the conventional results in the high $q^2$ region.

The expectation value is straightforward to compute in the limit where
the gluon field is treated as a classical background field.  If we write
\be
        < T(J^\mu (x) J^\nu(y))> = <T\left(\overline \psi (x) \gamma^\mu \psi(x)
\overline \psi(y) \gamma ^\nu \psi(y)\right) >\, ,
\ee
then when the background field is classical, and one ignores quantum
corrections arising from either loops of fermions or loops of gluons,
(a good approximation in the weak coupling limit of high parton 
densities), we obtain
\be
        <T(J^\mu(x) J^\nu(y))> = {\rm Tr}
(\gamma^\mu S_A(x)) {\rm Tr} (\gamma^\nu S_A(y)) +
{\rm Tr}(\gamma^\mu S_A(x,y) \gamma^\nu S_A(y,x) ) \, .
\label{tad}
\ee
In this expression, $S_A(x,y)$ is the Green's function for the fermion
field in the external field $A$
\be
        S_A(x,y) = -i< \psi(x) \overline \psi (y) >_A
\ee
for fixed $A$ (before averaging over A).

The first term on the right hand side of Eqn.~\ref{tad} is a tadpole
contribution which does not involve a non--zero imaginary part.  It
therefore does not contribute to $W^{\mu\nu}$.  We find then that
\be
        W^{\mu \nu}(q^2,p\cdot q) &=& {1\over 2\pi}\sigma {P^+ \over m} 
{\rm Im} \int dX^- d^4x \,e^{iq\cdot x}\,
\langle{\rm Tr}\Big( \gamma^\mu S_A (X^- + x/2,X^- - x/2)\nonumber \\
&\times& \gamma^\nu 
S_A(X^- - x/2,X^- + x/2)\Big)\rangle \, .
\ee
\subsection{The Light Cone Fock Distribution Function and Structure
Functions}
\vskip 0.15in

The expression we derived above for $W^{\munu}$ is
entirely general and makes no reference to the operator product
expansion.  In particular, it is relevant at the
small x values and moderate $q^2$ where the operator product expansion
is not reliable~\cite{AMueller1}.  At sufficiently high $q^2$ though
(and for massless quarks) it should agree with the usual leading twist
computation of
the structure functions.   The fact that we do not have a valid operator
product expansion forces us to distinguish between two quantities which
are identical in leading twist.  The first is the Fock space distribution of
partons within a hadron.  The second are the parton structure functions
which are measured in deep inelastic scattering.  
In our analysis, at high values of $q^2$, these
expressions are identical. At smaller values, say those values typical of
the intrinsic transverse momentum scale $\alpha_S \mu$,
they are no longer the same and must be differentiated between.

We will derive below an expression for the
light cone quark Fock distribution in terms of the propagator in light
cone quantization~\cite{KogutSoper, BjKogutSoper}. The quark Fock space
distribution
is then simply related to the structure function $F_2$ for $q^2 >>
\alpha_S^2 \mu^2 $.
In light cone quantization, only the two component spinor projection
$\psi_+$ is dynamical. 
(Note: notation and conventions are discussed in appendix A.)
The other two spinor components $\psi_-$ (recall
that $\psi = \psi_- +\psi_+$) are defined via the light cone constraint
relation defined below in Eq.~\ref{constraint}. The dynamical fermions
can
then be written in terms of creation and annihilation operators as
\be
\psi_+ = \int_{k^+>0} {d^3 k\over{2^{1/4} (2\pi)^3}} \sum_{s=\pm {1\over
2}} \left[ e^{ik\cdot x} w(s) b_s (k) + e^{-ik\cdot x}
w(-s) d_s^\dagger (k)
\right] \, .
\ee
Above $b_s (k)$ is a quark destruction operator and destroys a
quark with momentum $k$ while $d_s^\dagger (k)$ is an anti--quark
creation operator and creates an anti--quark with momentum k. Also above 
the unit spinors $w(s)$ are defined as 
\be
w({1\over 2}) =\left( \begin{array}{c}
0 \\
1\\
0\\
0\\ \end{array} \right)\,\, ; \,\,
w(-{1\over 2}) =\left( \begin{array}{c}
0 \\
0\\
1\\
0\\ \end{array} \right)\, ,
\ee
Note that since
\be
d^\dagger (\vec{k_t},k^+; x^+; +{1\over 2}) &=& b(-\vec{k_t},-k^+; x^+;
-{1\over 2}) \nonumber \\
d^\dagger (\vec{k_t},k^+; x^+; -{1\over 2}) &=& b(-\vec{k_t},-k^+; x^+;
+{1\over 2}) \nonumber \,\, ,
\ee
one can show that
\be
w(s) b_s (k) = 2^{1/4}\int d^3 x\,e^{-ik\cdot x}\, \psi_{+,s} (x) \, .
\ee
The light cone Fock distribution function is defined in terms of
the creation and annihilation operators as
\be
{dN\over {d^3k}} = {1\over {(2\pi)^3}}\sum_s
\left[b_s^\dagger
b_s + d_s^\dagger d_s\right] = {2\over {(2\pi)^3}}
\sum_s b_s^\dagger b_s \,\, .
\ee
We have assumed above that the sea is symmetric between quarks and
anti--quarks. Combining the two equations above, we get
\be
{dN\over{d^3 k}} = {2\sqrt{2}\over {(2\pi)^3}} \int d^3 x\, d^3\, y e^{ik\cdot
(x-y)}
\psi_{+}^\dagger (x)\psi_{+} (y) \, .
\ee
Using the light cone identity
\be
{\rm Tr}\left[ \gamma^+ \psi (x){\psibar}(y)
\right] = \sqrt{2}\,\psi_{+} (x) \psi_{+}^\dagger (y) \, ,
\ee
we obtain the following expression for the sea quark
distribution function
\be
{dN\over {d^3 k}} = {2 i\over {(2\pi)^3}}\int d^3 x\, d^3 y\,
e^{-ik\cdot (x-y)}\,{\rm Tr}\left[\gamma^+ S_A(x,y)\right] \,\,
\label{distprop}
\ee
where the fermion propagator $S_A(x,y)$ is the light cone time ordered
product $S_A(x,y) = -i <T(\psi(x){\psibar}(y))>$ in the background field
$A^\mu$. In our effective action approach, as discussed in sections 2
and 3.1, we can replace $S_A(x,y)\longrightarrow S_{A_{cl}}(x,y)$ to obtain
the sea quark distribution in the classical background field.

In a nice pedagogical paper, (see Ref.~\cite{Jaffe} and references
within),
Jaffe has shown that the Fock space distribution function can be
simply related to the {\it leading twist} 
structure function $F_2$ by the relation
\be
F_2(x,Q^2) = \int_0^{Q^2} dk_t^2 x {dN\over{dk_t^2 dx}}  \, .
\label{ltf2}
\ee
Actually, Jaffe's expression is defined as the sum of the quark and
anti--quark distributions. At small x, these are identical and the
resulting
factor of 2 is already included in our definition of the light cone
quark distribution function. In the
following section, we will compute the fermion Green's function in the
classical background field. We will then use it in Eq.~\ref{distprop}
and the
above equation to show that we do indeed recover the standard
perturbative
result for $F_2$. For heavy quarks and/or moderate $Q^2$, structure
functions should be computed by inserting the fermion Green's function 
in the definition for $W^{\mu \nu}$. We will see that in general the structure
functions and the Fock space distribution functions are not the same.

\section{The Fermion Green's Function in the Classical Background Field}
\vskip 0.15in

In this section we shall derive an expression for the fermion propagator
in the classical gluon background field described in section 2.2.
The field strength carried by
these classical gluons is highly singular, being peaked about the source
(corresponding to the parton current at x values larger than those in
the
field) localized at $x^-=0$. Away
from the source, the field strengths are zero and the gluon fields are
pure
gauges on both sides of $x^-=0$. The fermion wavefunction is obtained by
solving the Dirac equation in the background field on either side of the
source
and matching the solutions across the discontinuity at $x^-=0$. Once the
eigenfunctions are known, the fermion propagator can be constructed in
the
standard fashion. We begin this section with a discussion of the
notation
and conventions, proceed to write down the solution of the Dirac
equation and
finally, construct explicitly the fermion propagator in
the
classical gluon background field. This expression is formally exact and
is
valid to all orders in the source color charge density.

\subsection{The Dirac Equation in the Classical Background Field}
\vskip 0.15in

In order to compute the propagator for a spinor field in the
fundamental representation of the gauge group propagating in the
background gauge field
\be
        A^a_+ & =  & 0  \nonumber \\
         A^a_- & = & 0  \nonumber \\
         \tau \cdot A_t & = & \theta (x^-) \kappa_t (x_t) \, ,
\ee
where $\kappa_t(x_t), t=1,2$ is a two dimensional pure gauge,
\be
        \kappa_t(x_t) = -{1 \over {ig}} V(x_t) \nabla_t V^\dagger (x_t)
\, ,
\ee
we first need to solve the Dirac eigenvalue equation which can be
written as
\be
        \left\{ \vec{\alpha}_t \cdot(\vec{p}_t - g \vec{A}_t ) -
\sqrt{2}p^+ \alpha^- -\sqrt{2}p^-\alpha^+ +\beta M\right\} \psi_\lambda
(x) =
\lambda \psi_\lambda (x)
\ee
for the spinor field $\psi$ and a corresponding equation for $\overline
\psi
(x)$. The $\alpha$'s and $\beta$ above are defined in appendix A and
and $p^\mu = -i\partial_{\mu}$. For $x^- <0$, the solution is
trivial and
is just the free spinor plane wave solution. For $x^- >0$ the solution
is less
trivial and is given by the non--Abelian analogue of the `Baltz' ansatz
~\cite{Baltz}.
The full solution of the Dirac equation in the classical background
field is
\be
        \psi^{\alpha, s}_{\lambda q} (x)& = &
\theta (-x^-) e^{iq\cdot x} u^\alpha_{s,\lambda} (q)
+\theta (x^-) {1\over \sqrt{2}}\int {{d^2 p_t} ~\over {(2\pi )^2}}\,dz_t
(V(x_t)V^\dagger (z_t))^{\alpha\beta}
e^{ip_t\cdot x_t-iq^- x^+} \nonumber \\ &\times& e^{z_t\cdot (q_t-p_t)}
\exp\left(-i {{(p_t^2+M^2-\lambda) } \over {2p^-}}\,x^-\right)
\left\{1+{{(\alpha_t\cdot
p_t+ \beta M)}\over {\sqrt{2}q^-}}\right\} \alpha^-
u^{\beta,s}_{\lambda} (q)
\, .\nonumber\\
\label{wavefn}
\ee
Above, the superscripts $\alpha, \beta$ denote the color index in the
fundamental representation and $s$ the
spinor index. The elementary spinors are normalized as ${\bar
u}_{s,\lambda}
(q) u_{s^\prime,\lambda^\prime} (q) = 2M
\delta_{ss^\prime}\delta_{\lambda
\lambda^\prime}$ and summed over spins $(u{\bar u})_{\mu\nu} =
(M-q\!\!/)_
{\mu\nu}$.

The interested reader will notice that
the above equation is not continuous across $x^-=0$. This is because
though
$\psi_- (x)$ is continuous across $x^-=0$, $\psi_-$ is related to
$\psi_+$
via the light cone constraint equation
\be
\psi_+ = {1\over {\sqrt{2}q^-}} \left[\alpha_t \cdot ({1\over
i}\partial_t-
gA_t) +\beta M\right]\psi_- \, ,
\label{constraint}
\ee
which is discontinuous across $x^-=0$ by the same amount as in the
previous
equation.

\subsection{Computation of the Fermion Propagator}
\vskip 0.15in

Having obtained the eigenfunctions for the Dirac equation in the
classical
background field we are now in a position to compute the fermion
propagator
in the classical background field. This is given by the relation
\be
S(x,y) = \int {d^4 q \over {(2\pi)^4}} {1\over {q^2 + M^2
-i\varepsilon}}
\sum_{pol} \psi_{q} (x) {\bar\psi}_{q} (y) \, ,
\ee
after identifying $q^+ = (q_t^2+M^2-\lambda)/2q^-$.
It is straightforward to check from the above expression that
$(q\!\!/+M)S(x,y)=(2\pi)^4
\delta^{(4)}(x-y)$. Substituting the eigenfunctions from
Eq.~\ref{wavefn}
in the above, we have
\be
S(x,y)& = & \theta (-x^-) \theta (-y^-)
S_0 (x-y) + \theta (x^-) \theta (y^-) \left(V(x_t) S_0 (x-y)
V^\dagger (y_t)\right) \nonumber \\
&+ &
\int {{d^4q} \over {(2\pi )^4}} {1 \over {q^2 + M^2 -i\varepsilon}}
e^{iq\cdot (x-y)} \int {{d^2p_t} \over {(2\pi )^2}}
d^2z_t~\bigg\{  \theta (x^-)
\theta (-y^-) \nonumber \\
&\times &
e^{ip_t\cdot(x_t - z_t)}
e^{\left[ -i {{(p_t+q_t)^2-q_t^2} \over {2q^-}}x^- \right]}
\left(V(x_t) V^\dagger (z_t)\right){1\over 2q^-}(M-q\!\!/-p_t\!\!\!/)
\gamma^- (M- q\!\!/) \nonumber \\
&+ &
\theta (-x^-) \theta (y^-)\, e^{-ip_t\cdot (y_t- z_t)}
\exp{\left[i {{(p_t+q_t)^2-q_t^2} \over {2q^-}}y^- \right]}
\left(V(z_t) V^\dagger (y_t)\right) \nonumber \\
&\times& {1\over 2q^-}(M-q\!\!/)\gamma^-
(M-q\!\!/-p_t\!\!\!/) \bigg\}\, ,
\label{propagator}
\ee
where the free fermion Green's function is
\be
S_0 (x-y)=\int {d^4q\over {(2\pi)^4}}
e^{iq\cdot (x-y)}{(M-q\!\!/)\over{q^2+M^2-i\varepsilon}} \, .
\ee
The translational symmetry of the Green's function in the $x^-$
direction is
of course broken by the presence of the source at $x^-=0$. In the
absence of
the source of color charge, it may be confirmed that the free fermion
propagator is recovered by putting $V=I$, where I denotes the unit
matrix in
the fundamental representation.

The reader will note that the propagator between two points on the same
side of the source, for either $x^-,y^- <0$ or $x^-,y^->0$ is the free
propagator or a gauge transform of it. The only non--trivial
contribution
comes from the pieces connecting points on the opposite sides of the
source. The $\theta(x^-)\theta(-y^-)$ piece can be written more simply
as
\be
S(x,y) = -i\int d^4 z V(x_t)\, S_0(x,z) \gamma^- \delta(z^-) S_0 (z,y) \,
V^\dagger (z_t) \, .
\label{harart}
\ee
An analogous expression holds for the $\theta(-x^-)\theta(y^-)$ piece of
the propagator. A similar simple expression for the scalar quark
propagator
~\cite{MV2} was found recently by Hebecker and Weigert~\cite{HW} (see also 
the recent work of Balitskii~\cite{Ian}).
The only difference between the form of the above result and that for
scalar quarks is the $\gamma^-$ matrix present here due to the different
spinor structure and a partial derivative $\partial_{z^+}$ absent here
due
to the $1/2q^-$ factor in Eq.~\ref{propagator}.

If we define
\be
        G(x_t,x^-) = \theta(-x^-) + \theta(x^-) V(x_t)
\ee
which is the gauge transformation matrix which transforms the gluon
field at hand to a singular field which has only a plus component,
$A^{\prime \mu} = \delta^{\mu +}\alpha(x_t)$, we then see that our
propagator has the form
\be
        S_A(x,y) &  = & G(x)S_0(x-y)G^\dagger(y)
             -i  \int d^4 z G(x)\Bigg\{ \theta(x^-)\theta(-y^-)
                     ( V^\dagger(z_t)-1) - \nonumber \\
                     & & \theta(-x^-)\theta(y^-)(V(z_t)-1)\Bigg\}
G^\dagger(y)
               S_0(x-z) \gamma^- \delta(z^-) S_0 (z-y) \, .
\ee
This very simple form of the propagator is useful in the manipulations below. 

In fact the current-current correlation function is explicitly gauge
invariant.  We may therefore use the singular gauge form of the
propagator for computing the current-current correlation function
\be
        S_A^{sing}(x,y)&  = & S_0(x-y) -i \int d^4z 
\Bigg\{\theta(x^-)\theta(-y^-)
(V^\dagger(z_t)-1) - \nonumber \\
& &  \theta(-x^-)\theta(y^-)(V(z_t)-1) \Bigg\} S_0(x-z)\gamma^- \delta(z^-)
S_0(z-y)
\label{singprop}
\ee
A diagrammatic representation of the form of the propagator above is
shown in Fig.~1
In the expressions below for $W^{\mu \nu}$ we will drop the superscript
$sing$ and simply use the singular gauge expression for the propagator.

\begin{figure}[ht]
\begin{center}
\setlength\epsfxsize{6in}
\epsffile{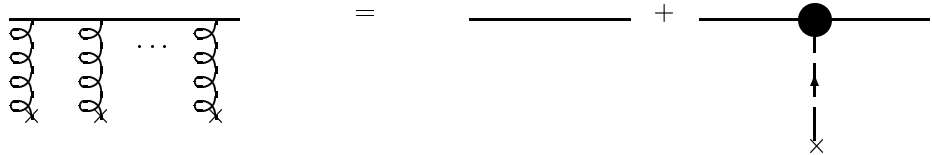}
\caption{Diagrammatic representation of the propagator in Eq.~48.}
\end{center}
\end{figure}

The Fock space distribution function however
is gauge dependent.  In computing it we must therefore either use the 
form of the propagator with the explicit gauge matrices above, or go 
back to our original form. In what follows, we shall use the original 
form of Eq.~\ref{propagator} for computing the Fock space distribution 
function.

Our result for the fermion propagator in the classical background
field was obtained for a $\delta$--function source in the $x^-$
direction. This assumption was motivated by the observation that small
x modes with wavelengths greater than $1/P^+$ perceive a source which
is a $\delta$-function in $x^-$. The propagator above can also be
derived for the general case where the source has a dependence on
$x^-$. The gauge transforms above are transformed from
$V(x_t)\rightarrow V(x_t,x^-)$, to path ordered exponentials, where
$V(x_t,x^-)$ is given by Eq.~\ref{puresoln}. Our result for the
propagator
is obtained as a smooth limit of $\Delta x^- = 1/xP^+ >> x^- (=1/P^+)$.
In other words, our form for the propagator is the correct one as long
as we interpret the $\theta$-functions and $\delta$-functions in $x^-$
to
be so only for distances of interest greater than $1/P^+$, the scale of
the classical source.

\section{The Leading Twist Computation of $F_2$}
\vskip 0.15in

Now that we have computed the fermion propagator in the classical
background
field, we are in a position to calculate the
sea quark distributions in this background field, and in turn the
structure function $F_2$ and evolution equations using Eq.~\ref{ltf2}.
This calculation is accurate
to lowest order in $\alpha_S$ but to all orders in $\alpha_S\mu$.
Due to the singular nature of the propagators in the background field,
the
actual computation of the distribution function is a little subtle and
will
be outlined below.

Before we go ahead to the computation, we will begin with a discussion
of
the averaging procedure over the labels of color charges at 
rapidities higher 
than those of interest. We will obtain a compact expression for it
below.

\subsection{Color Averaging Over the Sources of Color Charge at Higher
Rapidities}
\vskip 0.15in

Our expression for the propagator in the previous section makes no
particular assumption about the
color averaging over the color labels of the external sources
corresponding to
the valence quarks and/or gluons at higher rapidities than the rapidity
of
interest. The expression we
quote before averaging is the quantity which will be useful in loop
graph
computations.
To relate our expression to physical observables, as discussed in
section 2
(see Eq.~\ref{expvalue}) we need to average over all
the color labels of the external color charge density
corresponding to the color charge density
$\rho^a(x_t,y)$ at rapidities greater than the rapidity of interest.

Here we have smeared out the source in $x^-$ and are no longer treating
it as a delta function.  This means that the sources acquire a rapidity
dependence and that the weight for the Gaussian  fluctuations over
sources is replaced by
\be
        \int d^2 x_t dy {1 \over {2d\mu^2/dy}} \rho^2(x_t,y)
\ee
which leads us to define 
\be
        \mu^2 = \int_y^\infty dy^\prime \,{{d\mu^2} \over {dy^\prime}}
\ee
Here the lower limit would be the rapidity of interest for evaluating
the  structure functions.

If we average the Green's function in Eq.~\ref{propagator} over all
possible values of the color labels corresponding to the partons at
higher
rapidities, we can employ the following definitions for future
reference.
Defining
\be
{1\over N_c}\,<{\rm Tr}\left(V (x_t) V^\dagger (y_t)\right) >_{\rho} =  \gamma (x_t -
y_t) \, ,
\ee
we see that
\be
        \gamma (0) = 1 \, ,
\ee
which follows from the unitarity of the matrices $V$.
Now defining the Fourier transform~\footnote{We define the Fourier 
transform in this way because it corresponds to only the connected 
pieces in the correlator.} 
\be
        {\tilde\gamma} (p_t) = \int d^2x_t~ e^{-ip_tx_t} 
\left[\gamma(x_t)-1\right] \, ,
\label{Fourgamm}
\ee
we have the sum rule
\be
        \int {{d^2p_t} \over {(2\pi )^2}}~ {\tilde \gamma}(p_t) = 0 \, .
\label{rule}
\ee
The function ${\tilde \gamma} (p_t)$ will appear frequently in our
future
discussions and as we shall see, can be related to the gluon density at
small x.

For the particular case of a large nucleus~\cite{MV} the averaging
procedure has the form
\be
<O>_\rho = \int [d\rho] O(\rho)\exp\left(-\int_0^\infty dy \int d^2 x_t
{{{\rm Tr}\rho^2(x_t,y)} \over {d\mu^2(y)/dy}}\right) \, ,
\ee
where $\mu^2$ is the average color charge density per unit transverse
area
per unit rapidity. For an extensive
discussion of the above averaging procedure we refer the reader to
Refs.~\cite{MV,Kovchegov,JKLW}.
In appendix B we will explicitly derive an expression for
${\tilde \gamma}(p_t)$ for Gaussian fluctuations. A Gaussian form for
the
averaging over color charges at higher rapidities is likely valid for
very large nuclei or for realistic nuclei and hadrons for $x<<1$ but not
at too small $x$'s (or large enough parton densities) where non--linear
corrections to the renormalization group equations are important.

\subsection{Sea Quark Fock Space Distribution Function}
\vskip 0.15in

The relation Eq.~\ref{distprop} can now be combined with our expression
in
Eq.~\ref{propagator} to compute the sea quark distribution function. We
shall
use below the following identities:
\be
{\rm Tr}\left[\gamma^+ (M- q\!\!/)\right] &=& 4 q^+ \, \nonumber \\
{1\over 2q^-}{\rm Tr}\left[(M- q\!\!/-p_t\!\!\!/)\gamma^-
(M-q\!\!/)\gamma^+\right] &=& {2\over q^-}(M^2 + q_t^2 +p_t\cdot q_t)
\, .
\ee
We then obtain the following expression for the distribution function
\be
{dN^{ferm} \over {d^3k}} &=& {2i N_c  \over{(2\pi)^3}}
\int d^3 x\,d^3 y\, \int {d^4 q\over {(2\pi)^4}}\, {e^{i(q-k)\cdot
(x-y)}
\over {q^2+M^2-i\varepsilon}}\nonumber \\
&\times& \Bigg\{
4q^+\,\left[\theta(-x^-)\theta(-y^-)+\theta(x^-)\theta(y^-)
\gamma(x_t-y_t)\right]\nonumber \\
&+&\int {d^2 p_t\over{(2\pi)^2}}\,d^2 z_t\,{2\over
q^-}\left(M^2+p_t\cdot q_t
+q_t^2\right) \Bigg[\theta(x^-)\theta(-y^-)e^{ip_t\cdot(x_t -
z_t)}\nonumber
\\
&\times& \exp^{\left( -i {{(p_t+q_t)^2-q_t^2} \over {2q^-}}x^- \right)}
\gamma(x_t-z_t) +
\theta (-x^-) \theta (+y^-)\, e^{-ip_t\cdot (y_t- z_t)}\nonumber \\
&\times& \exp{\left(i {{(p_t+q_t)^2-q_t^2} \over {2q^-}}y^- \right)}
\gamma(z_t-y_t) \Bigg] \Bigg\} \, .
\label{sea2}
\ee
We will now sketch below the procedure used to simplify the above
equation.

a) First perform the integrals over the transverse co--ordinates. This
introduces a common factor $\pi R^2$ and for the
$\theta(-x^-)\theta(-y^-)$ term
a factor $\delta^{(2)}(q_t-k_t)$ (a factor ${\tilde \gamma}$ which was
defined in Eq.~\ref{Fourgamm} pops up in the other three pieces).
\vskip 0.075in

b) Perform the integrals over $x^-$ and $y^-$. For the $\theta(\pm x^-)
\theta(\pm y^-)$ pieces, we obtain the factors
\be
{1\over {q^+-k^+\pm i\varepsilon}}\,{1\over {q^+-{k^\prime}^+\mp
i\varepsilon}}
\, ,
\ee
respectively. We have introduced above (to ensure smooth convergence) a
slight difference in the momenta ($k^+$ and ${k^\prime}^+$ respectively)
multiplying $x^-$ and $y^-$ in the phases. In the final step we take
the limit ${k^\prime}^+-k^+\rightarrow 0$.
\vskip 0.075in

c) The simple contour integral over $q^+$ is done next. This introduces
the factors $\theta(\pm q^-)$.                                  

d) The final step is to perform the (logarithmic) integral over $q^-$.
The ultraviolet cutoff $\Lambda\rightarrow \infty$ cancels among the
different terms and we obtain a finite result. Putting all the
terms together and using the identity in Eq.~\ref{rule} , we obtain the
general result for the sea quark distribution
function
\be
{1\over {\pi R^2}}{{dN^{ferm}} \over {dk^+ d^2 k_t}} &=&
 {N_c\over {2\,\pi^4}}
{1\over k^+} \int {d^2p_t \over {(2\pi )^2}}\,\, {\tilde \gamma} (p_t)
\nonumber \\
&\times&\left[1- {{\left(k_t^2 + k_t\cdot p_t+ M^2\right)}
\over {p_t^2+2k_t\cdot p_t}}
\log\left({{(k_t+p_t)^2 + M^2}\over {k_t^2+M^2}}\right)\right]
.\nonumber\\
\label{seafinal}
\ee

In the region where the logarithm can be expanded, it can be checked
analytically that the argument of the above expression is positive
definite. We have checked numerically that the argument remains positive
definite in the entire $(k_t,p_t)$ phase space (as it should be).

In the next section we will study the above result in different limits
and
relate it to the well known evolution equations for sea quark
distributions.

\subsection{Evolution Equations for Sea Quark Distributions at Small x}
\vskip 0.15 in

In this section, we will show that for large $q^2$, 
the Fock space seaquark distribution 
we derived above in Eq.~\ref{seafinal} gives us the Altarelli--Parisi
evolution equation for seaquark evolution at small x.

Towards that end, consider the Fock space distribution in the 
limit of large $k_t$. In the integral in Eq.~\ref{seafinal}
we approximate $k_t>>p_t$. Then expanding the logarithm and 
defining 
\be
{{(k_t+p_t)^2-k_t^2}\over {k_t^2+m^2}} = 1 + \kappa\, ,
\ee
we find that the terms in the square brackets $[\cdots]$ in
Eq.~\ref{seafinal} can be approximated by
\be
\left[\cdots\right]\approx \Bigg[ {\kappa\over 2} -{\kappa^2\over 3} 
-{k_t\cdot p_t\over {k_t^2+M^2}} +{(k_t\cdot p_t)\kappa\over {k_t^2+M^2}}
-{\kappa^2\over 2}{(k_t\cdot p_t)\over {k_t^2+M^2}} \Bigg] \, .
\ee
Now ${\tilde \gamma}(p_t)$ has rotational symmetry in the transverse
plane. This helps simplify our expression above since only even terms in
$k_t\cdot p_t$ survive. To leading order in $p_t^2/k_t^2$ then,
Eq.~\ref{seafinal} reduces to
\be
{1\over {\pi R^2}}k^+{{dN^{ferm}} \over {dk^+ d^2 k_t}}={N_c\over
{2\pi^4}}
\int {{d^2 p_t}\over {(2\pi)^2}} {\tilde \gamma} (p_t) {1\over 2}
\left[{p_t^2\over {k_t^2+M^2}} -{4\over 3} {k_t^2 p_t^2\over
{(k_t^2+M^2)^2}} + {k_t^2 p_t^2\over{(k_t^2+M^2)^2}} 
\right] \, .\nonumber\\
\ee
Since we are interested in the limit $k_t^2>>M^2$, the above expression can
be further simplified to read
\be
{1\over {\pi R^2}}k^+{{dN^{ferm}} \over {dk^+ d^2 k_t}}={N_c\over
{4\pi^4}} {2\over 3}
{1\over k_t^2}\int {{d^2 p_t}\over {(2\pi)^2}}
p_t^2\,{\tilde \gamma} (p_t) \, .
\label{seasimple}
\ee

To make contact with the evolution equations we will now obtain a
relation
between ${\tilde \gamma}(p_t)$ and the gluon distribution function at
small x. We begin with the relation we defined in the last section--Eq.
~\ref{Fourgamm}:
\be
        {\tilde\gamma} (p_t) = \int d^2x_t~ e^{-ip_tx_t}
[\gamma(x_t)-1] \, .
\nonumber
\ee
Then
\be
p_t^2 {\tilde\gamma} (p_t) = -\int d^2 x_t\, e^{ip_t\cdot x_t}\, 
\partial_{x_t}^2 \gamma(x_t)
\, .
\label{gamm1}
\ee

Recall that $\gamma(x_t) = {1\over N_c} <{\rm Tr} (V(x_t)V(0)^\dagger>_\rho$. 
Expanding out the matrix $V = 1 + i\Lambda(x_t)
-
  {\Lambda^2 }(x_t)/2 + \cdots $'', and doing the same for the
correlator of
gauge fields $A^i=\frac{-1}{ig}V\partial^i V^\dagger$, we obtain the
relation
\be
p_t^2 {\tilde\gamma}(p_t) = {g^2 \over 2N_c}\int d^2 x_t\, e^{ip_t\cdot x_t} 
<A_i^a (x_t) A_i^a (0)>_\rho \, .
\label{gammA2}
\ee

The correlator $<A_i^a A_i^a>$ can be related to the gluon distribution
function by the formula
\be
{dN\over {d^3 l}} = {2 \mid l^+\mid \over {(2\pi)^3}}\int d^3 x d^3 {\bar
x^\prime} \,e^{-il^+ x^-} \,e^{+i l^+ {x^\prime}^-}\, 
\theta(x^-)\theta({x^\prime}^-) 
\,e^{il_t\cdot x_t } <A_i^a (x_t) A_i^a (0)>_\rho \, .
\label{correlator}
\ee

Integrating both sides over $l^+$, we obtain
\be
\int d^2 x_t e^{il_t\cdot x_t} \langle A_i^a (x_t) A_i^a (0) \rangle =
{k^+\over {\pi R^2}} {(2\pi)^3 \over 2} \int_{k^+}^{P^+} {dl^+ \over 
\mid l^+ \mid} {dN^{glue}\over {d^2 l_t dl^+}} \, .
\ee
Substituting the RHS of the above equation in Eq.~\ref{gammA2} and the 
resulting expression for $p_t^2{\tilde \gamma}(p_t)$ into
Eq.~\ref{seasimple}, we obtain
\be
        x {{dN^{ferm}} \over {dxdk^2_t}} = {\alpha_S \over {2\pi}}
{2\over 3} {1\over k_t^2}
 \int_x^1 {dy \over y} \int_0^{k_t^2} 
dp_t^2 x {{dN^{glue}} \over {dy dp_t^2}} \, ,
\ee
where we defined $x=k^+/P^+$ and $y = l^+/P^+$. 

Recall that the structure function to leading twist and lowest order in
$\alpha_S$ is
\be
F_2(x,Q^2) = 
\int_0^{Q^2} dk_t^2 x{dN^{ferm}\over {dx dk_t^2}} = x[q(x,Q^2) + {\bar
  q}(x,Q^2)] \equiv 2xq(x,Q^2)\, , 
\ee
and similarly,
\be
xG (x,Q^2) = \int_0^{Q^2} x {dN^{glue}\over {dx dk_t^2}}
\, ,
\ee
where $xq(x,Q^2)$ and $x G(x,Q^2)$ are the quark and gluon momentum
distributions respectively. At small x, the sea is symmetric and we
take $q(x,Q^2) = {\bar q} (x,Q^2)$. 
In the limit of light quark masses
$Q^2>>M^2$
we find then that
\be
{d(xq(x,Q^2))\over {d\log (Q^2)}} = {\alpha_S\over {4\pi}}\,{2\over 3} x \int_x^1
{dy \over y^2} yG(y,Q^2) \, .
\ee
Now at small x, $yG(y,Q^2)$ is slowly varying. One can for instance 
parametrize it by
a power law $1\over x^{C\alpha_S}$, where C is some constant. The scale of
variation of the structure function then corresponds to higher orders in
$\alpha_S$. We can therefore take $yG(y,Q^2)$ out of the integrand. (The same
is true for any other slowly varying function~\cite{GoussPirn}.)  We 
then get finally for the seaquark evolution equation at small x the 
result
\be
{d(xq(x,Q^2))\over {d\log (Q^2)}} = {\alpha_S\over {4\pi}}\,{2\over 3} 
xG(x,Q^2) + O(\alpha_S)\, .
\label{Fockevolve}
\ee
Thus to lowest order in $\alpha_S$, the seaquark evolution at small x is
local and simply proportional to the gluon density at that x.

Now consider the Altarelli--Parisi evolution equation
\be
Q^2 {d\Sigma\over dQ^2} = {\alpha_S(Q^2)\over {2\pi}} \left[\Sigma \otimes
P_{qq} + G \otimes 2fP_{qG}\right] \, .
\ee
Above the operation $\otimes$ denotes
\be
A\otimes B\equiv \int_x^1 {dy\over y} A(y) B({x\over y}) \, ,\nonumber
\ee
and $\Sigma = \sum_f (q + {\bar q})$, where $f$ is the number of flavors.
Also, $P_{qq}$ and $P_{qG}$ are the well known Altarelli--Parisi 
splitting functions. 
Since at small x the quark distribution is $\alpha_S$ suppressed relative to 
the glue, taking $q={\bar q}$ and $f=1$ the leading contribution to
the seaquark evolution is
\be
{d(xq(x,Q^2))\over {d\log (Q^2)}} = {1\over 2}\,{\alpha_S\over {2\pi}}\,
\int_x^1 {dy\over y} G(y,Q^2)\Big[{x^2\over y^2} + \big(1-{x\over y}\big)^2
\Big]\, .
\ee
Above we have made use of the relation $2f P_{qG} (z) = f [z^2 + (1-z)^2]$.
Let $z=y/x$. Then the above relation can be re--written as
\be
{d(xq(x,Q^2))\over {d\log (Q^2)}} = {\alpha_S\over {4\pi}}\,
\int_1^{1\over x} {dz\over z^2} zG(zx,Q^2)\Big[{2\over z^2} + 1-
{2\over z}\Big] \, .
\ee
Again, as previously, we can argue that since at small x $zG(zx,Q^2)$ is
slowly varying, we can take it out of the integral. Doing that and 
performing the integral, we obtain finally
\be
{d(xq(x,Q^2))\over {d\log (Q^2)}} = {\alpha_S\over {4\pi}}\,{2\over 3} 
xG(x,Q^2) + O(\alpha_S)\, .
\ee
which is the same as Eq.~\ref{Fockevolve}. This form of the sea quark 
evolution equation at small x was first obtained by Ellis, Kunzt and 
Levin~\cite{EKL}.

\section{Computation of Current-Current Correlator to All Twists in
the Classical Background Field at small x.}
\vskip 0.15in

In the previous section we used the fermion Green's function derived in
section 4.2 to compute the lightcone seaquark distributions at small x
and subsequently the leading twist expression for the structure
functions.
It was shown that these structure functions obeyed evolution equations
which
were precisely the small x Altarelli--Parisi evolution equations.

In this section we will again use the fermion Green's function in
Eq.~\ref{propagator} to derive
an explicit result for the hadronic tensor $W^{\munu}$. This result will
be valid to all twists at small x and for arbitrary quark masses. For
light
quarks, we will
compute the structure functions $F_1$ and $F_2$ and show that the
leading
twist result in the previous section is recovered
as a limit of our general result. We will also use our general result to
obtain expressions for heavy quark structure functions at small x. We should 
note here that Levin and collaborators have studied screening corrections 
to the structure functions for light and heavy quarks 
in the Glauber--Gribov framework~\cite{Levin2}.

\subsection{Analytic Result for $W^{\munu}$ at Small x}
\vskip 0.15in

As in the previous sections, we define
\be
W^{\munu} (q, P, X^-) = {\rm Im}\, \int d^4 z\, e^{iq\cdot z}\,
<T(J^\mu (X^- + {z\over 2})J^\nu (X^- - {z\over 2})> \, ,
\ee
where ``Imaginary'' stands for the discontinuity in $q^-$. Then
\be
& &W^{\munu} (q, P) = {1\over 2\pi}\sigma {P^+ \over m} 
\int dX^- \, W^{\munu} (q, P, X^-)
\equiv {1\over 2\pi}\sigma P^+ {\rm Im}\,\int dX^- \int d^4 z\, e^{iq\cdot z} \nonumber \\
&\times& {\rm Tr}\left(
S_{A_{cl}}\left(X^- + {z\over 2}, X^- -{z\over 2}\right)\,\gamma^\nu
S_{A_{cl}}\left(X^- - {z\over 2}, X^- +{z\over
2}\right)\,\gamma^\mu\right) \, .
\ee
The only terms in the propagator that contribute to the above are
the $\theta(\pm x^-)\theta(\mp y^-)$ pieces. Using our representation
for
the propagator in Eq.~\ref{harart},
\be
& &W^{\munu}(q,P,X) = 
{\rm Im} N_c\,\int d^4 z\, e^{iq\cdot z} \int {d^4 p\,
d^4 l
\over
{(2\pi)^8}}\,{d^4 p^\prime \, d^4 l^\prime \over {(2\pi)^8}}\,d^2 u_t
d^2 u_t^\prime (2\pi)\delta(p^- - l^-)\nonumber \\
&\times&(2\pi)
\delta({p^\prime}^- - {l^\prime}^-) e^{i p\cdot (X + {z\over 2}-u_t)}
\, e^{-il\cdot (X-{z\over 2} -u_t)}\, e^{l^\prime\cdot (X-{z\over 2}-
{u^\prime)}_t}\, e^{-ip^\prime\cdot (X+ {z\over 2} -u_t^\prime)}
\nonumber \\
&\times& {\rm Tr} \left\{ (M-p\!\!/)\gamma^- (M-l\!\!/)\gamma^\mu
(M-{l^\prime}\!\!/)\gamma^- (M-{p^\prime}\!\!/)\gamma^\nu \over
{ (p^2 + M^2 -i\varepsilon)\,(l^2 + M^2 -i\varepsilon)\,
({l^\prime}^2 + M^2 -i\varepsilon)\,({p^\prime}^2 + M^2 -i\varepsilon)}
\right\}
\nonumber \\
&\times& \left( \theta(X^- + {z\over 2})\theta({z\over 2}-X^-) +
\theta(X^- -{z\over 2})\theta(-X^- - {z\over 2}\right)
\gamma(u_t-u_t^\prime)
\, .
\label{expwmunu}
\ee
We have used above the condition
$\gamma(u_t,u^\prime_t) =
\gamma(u_t^\prime-u_t)=\gamma(u_t-u_t^\prime)$, since the correlation
functions are translationally, rotationally and parity invariant in the
transverse plane. We now
perform the integral over $X^-$ and use the identity
\be
\int dX^- & &\!\!\!\!\!\!\!\!\!\!\!\!
e^{i(-p^+ + l^+ -{l^\prime}^+ + {p^\prime}^+)X^-}
\left(\theta(X^- + {z\over 2})\theta({z\over 2}-X^-) +
\theta(X^- -{z\over 2})\theta(-X^- - {z\over 2}\right) \nonumber \\
&=& \epsilon (z^-) {1\over {(p^+ - l^+ - {p^\prime}^+ + {l^\prime}^+)}}
2\sin\left({z^-\over 2} (p^+ - l^+ + {l^\prime}^+ -{p^\prime}^+)\right)
\, .
\ee
We then obtain
\be
 & &W^{\munu}(q,P) = {{\sigma P^+ N_c}\over {2\pi m}}  
{\rm Im} \int d^4 z e^{iq\cdot z} \epsilon(z^-)
\int {d^4 p d^4 l \over
{(2\pi)^8}}\,{d^4 p^\prime d^4 l^\prime \over {(2\pi)^8}}\,(2\pi)^2
\delta(p^- - l^-)\delta({p^\prime}^- - {l^\prime}^-) \nonumber \\
&\times&
(2\pi)^2 \delta^{(2)} (l_t-p_t+p_t^\prime-l_t^\prime)\,{\tilde\gamma}(
{l_t-p_t-p_t^\prime+l_t^\prime\over 2})\,
e^{i z_t\cdot (p_t + l_t -p_t^\prime -l_t^\prime)/2}\,
e^{iz^+ (p^- - {p^\prime}^-)}\nonumber \\
&\times&{\rm Tr} \left\{ (M-p\!\!/)\gamma^- (M-l\!\!/)\gamma^\mu
(M-{l^\prime}\!\!/)\gamma^- (M-{p^\prime}\!\!/)\gamma^\nu \over
{ (p^2 + M^2 -i\varepsilon)\,(l^2 + M^2 -i\varepsilon)\,
({l^\prime}^2 + M^2 -i\varepsilon)\,({p^\prime}^2 + M^2 -i\varepsilon)}
\right\}\nonumber \\
&\times& {1\over{(p^+ - l^+ - {p^\prime}^+ + {l^\prime}^+)}}
2\sin\left({z^-\over 2} (p^+ - l^+ + {l^\prime}^+ -{p^\prime}^+)\right)
\, .
\ee

The subsequent procedure of solving the integrals is as follows:
\vskip 0.1in

\noindent
a) Perform first the integral over $l_t^\prime$. Then perform the
integral
over $z_t$ and $z^+$. Defining $k_t = (p_t-l_t+p_t^\prime-l^\prime)/2$,
this sets $l_t^\prime = p_t-k_t-q_t$, $l_t = p_t-k_t$ and
$q_t=p_t-p_t^\prime$. Also, we get $q^- = p^- - {p^\prime}^-$.
\vskip 0.1in
\noindent
b) Next perform the integral over $z^-$. This gives us
\be
& &\int dz^- \epsilon(z^-)\, e^{iq^+ z^-} \Bigg( e^{-iz^-(p^+
-{p^\prime}^+)}
-e^{-iz^- (l^+ - {l^\prime}^+)} \Bigg)\nonumber \\
&=& i\Bigg[ {{l^\prime}^+ -
l^+ - {p^\prime}^+ + p^+\over {(q^+ - p^+ + {p^\prime}^+ + i\varepsilon)
(q^+ - l^+ + {l^\prime}^+ + i\varepsilon)}}
+ (i\varepsilon \rightarrow -i\varepsilon) \Bigg] \, .
\ee
The numerator above cancels the term $1/(p^+ - l^+
-{p^\prime}^+ +{l^\prime}^+)$ in $W^{\munu}$.
\vskip 0.1in
\noindent
c) Lastly, we do the integrals over ${p^\prime}^+$ and ${l^\prime}^+$.
This sets ${p^\prime}^+ = p^+ -q^+$ and ${l^\prime}^+ = l^+ - q^+$. Then
we can define in $W^{\munu}$, ${p^\prime} = p-q$ and ${l^\prime}= l-q$
with $l= p-k$ and $k^-=0$.

After these considerations, we can write $W^{\munu}$ as
\be
W^{\munu}(q,P) &=& {{\sigma P^+ N_c} \over {2\pi m}} {\rm Im} \int {d^4 p
\over
{(2\pi)^4}}\,{d^2 k_t \over {(2\pi)^2}}{dk^+\over {(2\pi)}}
\,\,{\tilde\gamma}(k_t) \nonumber \\
&\times&{\rm Tr} \left\{ (M-p\!\!/)\gamma^- (M-l\!\!/)\gamma^\mu
(M-{l^\prime}\!\!/)\gamma^- (M-{p^\prime}\!\!/)\gamma^\nu \over
{ (p^2 + M^2 -i\varepsilon)\,(l^2 + M^2 -i\varepsilon)\,
({l^\prime}^2 + M^2 -i\varepsilon)\,({p^\prime}^2 + M^2 -i\varepsilon)}
\right\}\, , \nonumber \\
\label{wmunu}
\ee
where $l^\prime$, $p^\prime$ and $l$ are defined as in step `c' above.
Correspondingly,  we can write $W^{\munu}$ as the imaginary
part of the diagram shown in Fig.~2.

\begin{figure}[ht]
\begin{center}
\setlength\epsfxsize{4in}
\epsffile{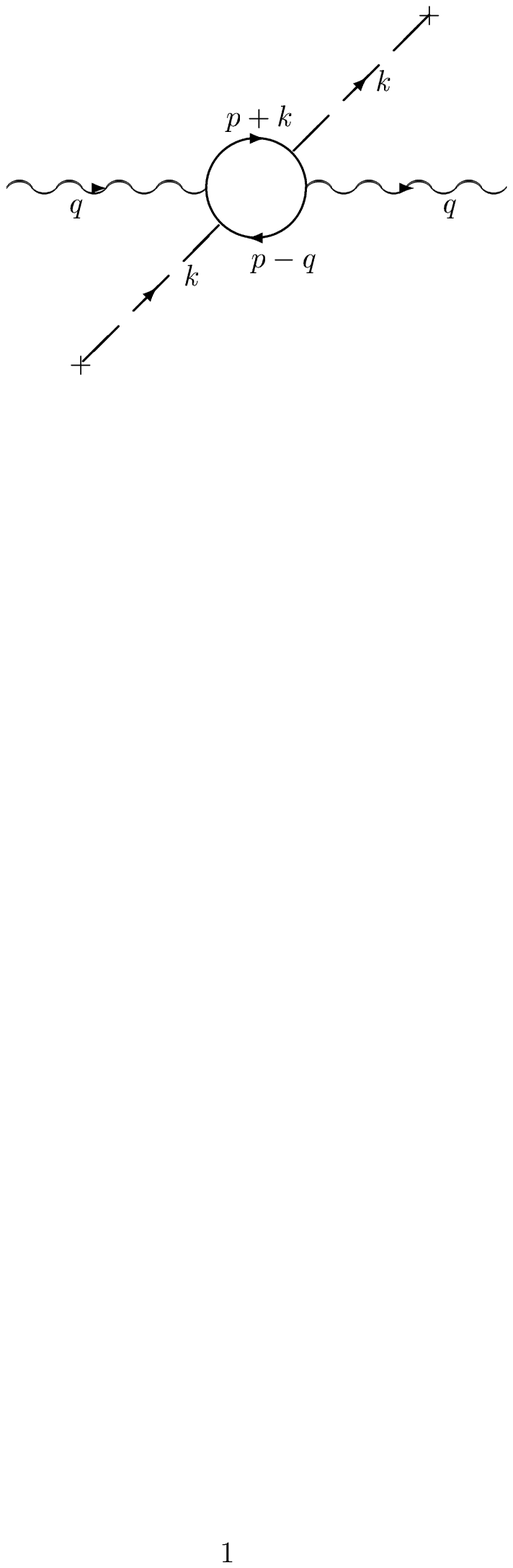}
\caption{Polarization tensor with arbitrary number of insertions from the
classical background field. The wavy lines are photon lines, the solid
circle denotes the fermion look and the dashed lines are the insertions
from the background field (see Fig.~1). The imaginary part of this diagram
gives $W^{\munu}$.}
\end{center}
\end{figure}

For the case of deep inelastic scattering, $q^2 > 0$ (see footnote~1), 
and we can cut the
above diagram only in the two ways shown in Fig.~3 (the diagram where both
insertions from the external field are on the same side of the cut
is forbidden by the kinematics).

\begin{figure}[ht]
\setlength\epsfxsize{6in}
\epsffile{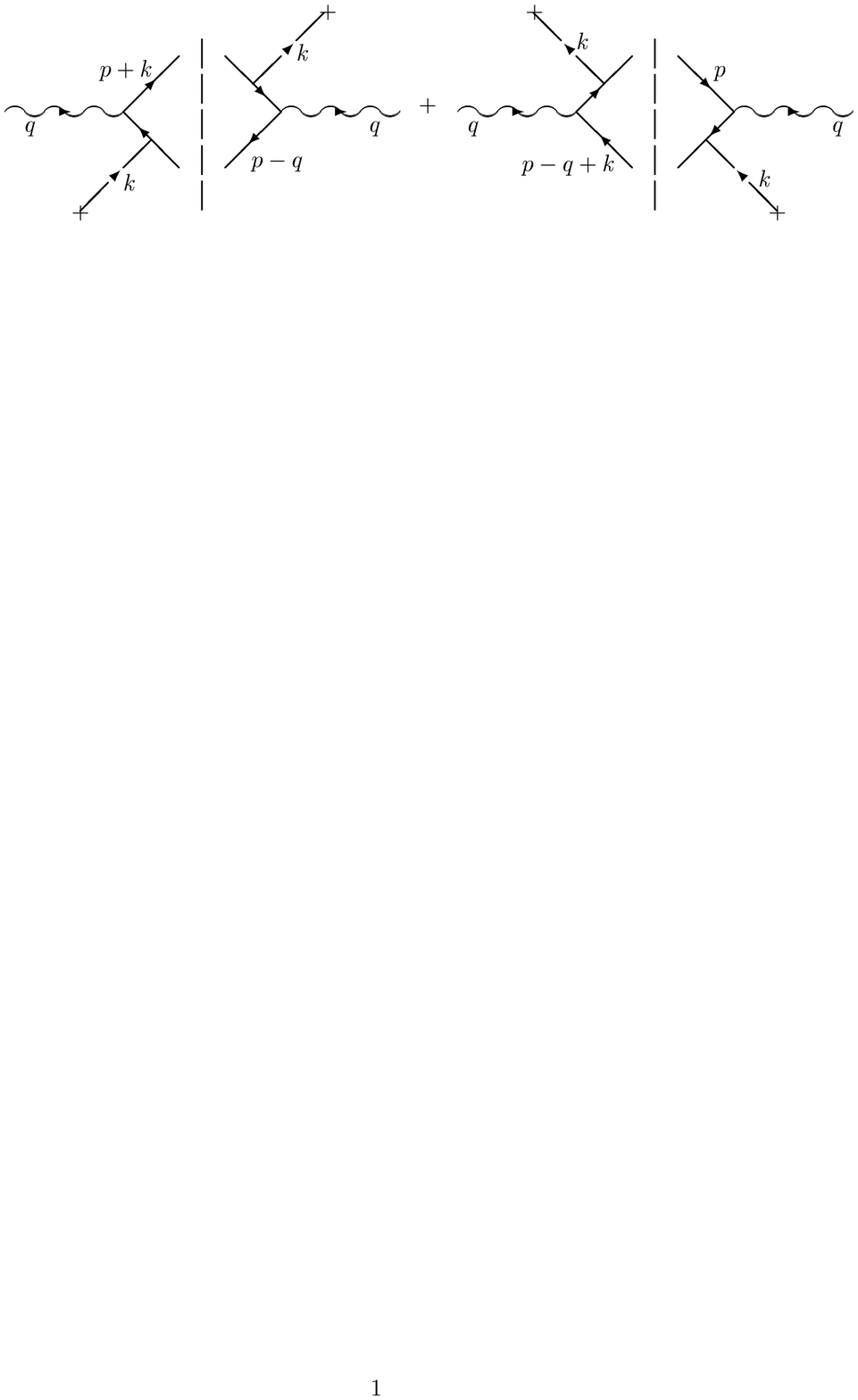}
\caption{Cut diagrams corresponding to the imaginary part of $W^{\munu}$.}
\end{figure}

Also interestingly, the contribution to $W^{\munu}$ can be represented
solely
by the diagram in Fig.~4 and not, as is usually the case, from the sum of
this
diagram and the standard box diagram. This is because in our
representation
of the propagator multiple insertions from the external field on a quark
line
can be summarized into a single insertion. See for instance Eq.~\ref{singprop}
which makes this point clear.

\begin{figure}[ht]
\setlength\epsfxsize{5.5in}
\epsffile{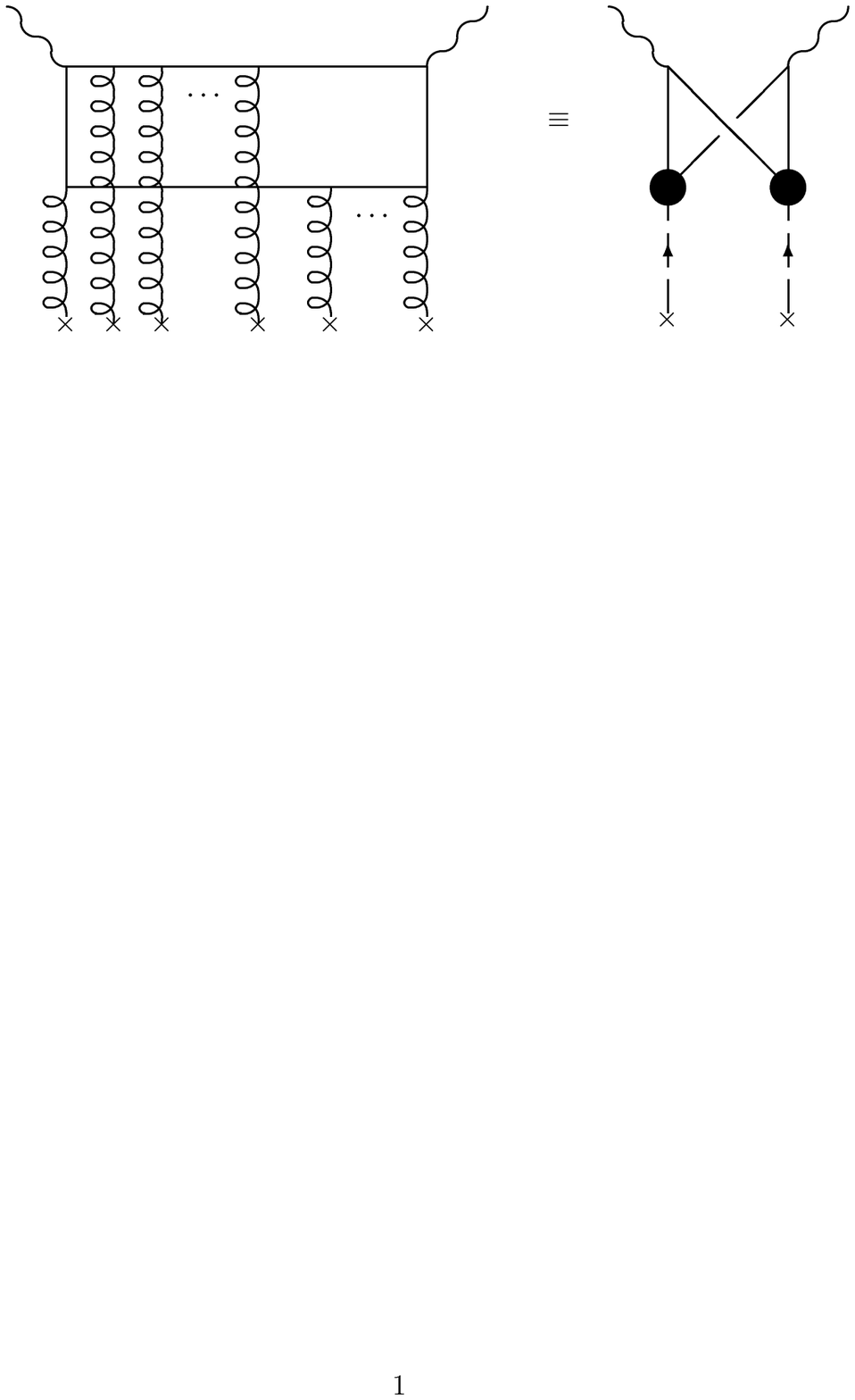}
\caption{In the singular gauge representation for the propagator (see 
Eq.~\ref{singprop} and Fig.~1), multiple, higher twist contributions from
the classical gluon background field to the current--current correlator
(imaginary part of LHS) is equivalent to the imaginary part of RHS.}
\end{figure}

Applying the Landau--Cutkosky rule, and making the shift $p\rightarrow
p+k$, Eq.~\ref{wmunu} can be written as
\be
& &W^{\munu}(q,P) = {{\sigma P^+ N_c} \over {2\pi m}}\int {d^4 p \over
{(2\pi)^4}}\,{d^2 k_t \over {(2\pi)^2}}{dk^+\over {(2\pi)}}
\,\,{\tilde\gamma}(k_t)\nonumber \\
&\times& {\rm Tr}
\left\{ (M-p\!\!/-k\!\!/)\gamma^- (M-p\!\!/)\gamma^\mu
(M-p\!\!/+q\!\!/)\gamma^- (M-p\!\!/-k\!\!/+q\!\!/)\gamma^\nu
\right\}\nonumber \\
&\times& \Bigg[ \theta(p^+ + k^+)\theta(q^+ -p^+) (2\pi)^2
\delta((p+k)^2 + M^2)\,\delta((p-q)^2 + M^2))
{1\over {p^2 + M^2}}\nonumber \\
&\times& {1\over {(p+k-q)^2 + M^2}}
+ \theta(p^+)\theta(q^+ - p^+)(2\pi)^2
\delta(p^2 + M^2)\,\delta((p+k-q)^2 + M^2)) \nonumber \\
&\times&{1\over {(p+k)^2 + M^2}}\,{1\over {(p-q)^2 + M^2}}\Bigg]\, .
\ee
With an appropriate change of variables, the second term is the same as
the first except that now $\mu \leftrightarrow \nu$. We then get
\be
& &W^{\munu}(q,P) = {\sigma P^+ N_c\over {2\pi m}}\int {d^4 p \over
{(2\pi)^4}}\,{d^2 k_t \over {(2\pi)^2}}{dk^+\over {(2\pi)}}
\,\,{\tilde\gamma}(k_t) M^{\munu}\theta(p^+ + k^+)\theta(-p^+)\nonumber
\\
&\times& (2\pi)^2
\delta((p+k)^2 + M^2)\,\delta((p-q)^2 + M^2))
{1\over {p^2 + M^2}}\,{1\over {(p+k-q)^2 + M^2}}\, ,\nonumber\\
\ee
where above the trace is represented by~\footnote{Kinematic note: the 
observant reader will notice we have put $q^+=0$ here. Since 
we are working in the infinite momentum frame, the hadron has only one 
large momentum component, $P^+$. The rest are put to zero. For the photon, 
we choose a left moving frame such that $q^0 = |q^z|$ and $q^+=0$. Then, 
$q^2 = q_t^2 >0$, $P\cdot q= -P^+ q^-$ and $x_{Bj} = -q^2/(2P\cdot q)
\equiv q_t^2/(2P^+ q^-)$. Since in the infinite momentum frame 
$ 0<x_{Bj}<1 $, this gives $ q^- > 0$. We are at liberty to choose the above 
frame since the hadron tensor is clearly Lorentz invariant and hence 
can be expressed purely as a function of $q^2$ and $P\cdot q$. The 
explicit presence of $P^+$ in Eq.~\ref{expwmunu} may give the reader cause
for concern. It arises from a relativistic normalization of the vacuum
states. One can show that, despite appearances, Eq.~\ref{expwmunu} is
Lorentz invariant. Of course our later results will confirm this fact.} 
\be
M^{\munu} = {\rm Tr}
\Bigg\{ (M-p\!\!/-k\!\!/)\gamma^- (M-p\!\!/)\gamma^\mu
(M-p\!\!/+q\!\!/)\gamma^- (M-p\!\!/-k\!\!/+q\!\!/)\gamma^\nu
+ \mu\leftrightarrow \nu \Bigg\} \, .\nonumber\\
\ee
Performing the $\delta$--function integrations above, our expression for
$W^{\munu}$ can be simplified to
\be
W^{\munu} = {-\sigma P^+ N_c\over {16\pi m}}\,{1\over {(q^-)^2}}\,\int
{d^2 k_t \over {(2\pi)^2}}\,{d^2 p_t \over
{(2\pi)^2}}\,{\tilde\gamma}(k_t)
\int_{-\infty}^{-M_{p-q}^2\over {2q^-}} {dp^+\over 2\pi}
{M^{\munu}\over {M_{p+k-q}^2}}\cdot I(k_t,p_t,q,p^+) \, ,
\label{finwmunu}
\ee
with the definitions $M_{p-q}^2 = (p_t-q_t)^2 + M^2$ and
$M_{p+k-q}^2 = (p_t+k_t-q_t)^2 + M^2$, and
\be
I(p_t,k_t,q,p^+) =
{1\over {p^+ - {(M_p^2 - M_{p-q}^2)\over {2q^-}}}}\,\,\,\,
{1\over {p^+ - {M_{p-q}^2(M_{p+k}^2-M_{p+k-q}^2)\over {(2q^-)M_{p+k-q}^2
}}}}
\, .
\ee
Eq.~\ref{finwmunu} is our general result for the hadronic tensor. We
shall
now study the different components of the above tensor and extract from
these, different limits of interest to us.

\subsection{Structure Functions at Small x}
\vskip 0.15in

The hadronic tensor $W^{\munu}$ can be decomposed in terms of the
structure
functions $F_1$ and $F_2$ as~\cite{Pokorski}
\be
m W^{\munu} &=& -\Bigg(g^{\munu} - {q^\mu q^\nu \over {q^2}}\Bigg) F_1
\nonumber \\
&+& \Big(P^\mu - {q^\mu (P\cdot q)\over {q^2}}\Big)\,\Big(P^\nu -
{q^\nu {P\cdot q}\over {q^2}}\Big) {F_2\over {(P\cdot q)}} \, , \label{strctfn}
\ee
where $P^\mu$ is the four--momentum of the hadron or nucleus and
$P^2= m^2 \approx
0$ ($<<q^2$). In the infinite momentum frame, we have $P^+\rightarrow
\infty$ and $P^-, P_t \approx 0$.
The above equation can be inverted to obtain expressions for $F_1$
and $F_2$ in terms of components of $W^{\munu}$. 
Since in our kinematics $q^+=0$ (see footnote 3 for a kinematic note) 
\be
F_1 = {F_2\over 2x} + \left({q^2 \over {(q^-)}^2}\right)\, W^{--} \, ,
\ee
with
\be
{1\over 2x} F_2 = -\left( {(q^-)^2 \over q^2}\right)\, W^{++} \, .
\ee

It is useful to verify explicitly that our expression for $W^{\mu \nu}$
derived in an external field can be written in the form of Eqn. \ref{strctfn}.
Recall that $W^{\mu \nu}$ can be written in Lorentz covariant form 
by using the vector $n^\mu = \delta^{\mu +}$.  Using $n \cdot \gamma =
-\gamma^-$ in Eqn. \ref{wmunu}, we see that $W^{\mu \nu}$ is a Lorentz 
covariant function of the only vectors in the problem--$q^\mu$ and $n^\mu$.
Identifying $n^\mu = P^\mu /P^+$ in Eqn. \ref{strctfn}, we see that these
forms are in complete agreement.  We also see that all factors of $m$
disappear from $F_1$ and $F_2$ by the explict forms of Eqns. \ref{strctfn}
and \ref{wmunu}.  Henceforth we will take $m=1$ since it disappears from the
quantites of interest, and was in fact only introduced due to historical
normalization conventions.  

We also see that the structure functions can only be functions of $q^2$
and $n\cdot q$ by Lorentz invariance.  We can therefore take $q^+ = 0$
for the purpose of computing $F_1$ and $F_2$.

To compute $W^{++}$ and $W^{--}$, we need to know the
the traces $M^{++}$ and $M^{--}$, respectively in Eq.~\ref{finwmunu}.
We can compute them explicitly and the results can be represented 
compactly as
\be
{1\over 16} M^{++} = {1\over 2}\left( M_p^2 M_{p+k-q}^2 + M_{p+k}^2 
M_{p-q}^2 - q_t^2 k_t^2 \right) \, ,
\ee      
and
\be
M^{--} = 32 (p^-)^2 (p^- - q^-)^2 \, . 
\ee
From the relations above of $F_1$ and $F_2$ to $W^{++}$ and $W^{--}$, 
we obtain from Eq.~\ref{finwmunu} the following general results for 
the structure functions for arbitrary values of $Q^2$, $M^2$ and the
intrinsic scale $\mu$,~\footnote{which is implicitly contained in the
function ${\tilde \gamma(k_t)}$ in Eq.~\ref{finwmunu}.}
\be
F_2 = {\sigma N_c\over 16\pi^2} \int {d^2 p_t \over {(2\pi)^2}} 
{d^2 k_t\over {(2\pi)^2}} {\tilde \gamma(k_t)} {M^{++}\over 
{\left(M_{p-q}^2 M_{p+k}^2 - M_{p+k-q}^2 M_p^2\right)}} \log\left(
{M_{p+k}^2 M_{p-q}^2 \over {M_{p+k-q}^2 M_p^2}}\right) \, ,
\label{generalf2}
\ee
and 
\be
F_1 = {F_2\over 2x} - {q^4 \sigma N_c\over {2x 2\pi^4}}\int d^2 p_t 
\int {d^2 k_t \over {(2\pi)^2}} {\tilde \gamma}(k_t){ {\cal F} (\alpha, 
\beta)\over M_{p-q}^2 M_{p+k-q}^2} \, .
\label{generalf1}
\ee
Above, we defined the function ${\cal F}$ in terms of $\alpha$ and $\beta$
(in turn functions of $p_t, k_t, q_t$ and $M$) which are defined as 
\be
\alpha = \left(1-{M_{p+k}^2\over M_{p+k-q}^2}\right) \,\, ; \,\,
\beta = \left(1-{M_p^2\over M_{p-q}^2}\right) \, .
\ee   
The general expression for ${\cal F}$ is
\be
{\cal F} (\alpha,\beta) &=& {1\over \alpha\beta} \Bigg[ {1\over 3} 
- {3\over 2}\big({1\over \alpha}+ {1\over \beta}\big) + 
\big({1\over \alpha^2} + {1\over \alpha\beta} + {1\over \beta^2}\big)
\nonumber \\
&+& {\alpha\beta\over {2(\beta-\alpha)}}\Big\{ {1\over \alpha^2}
\big(1-{1\over \alpha}\big)^2 \log(1-\alpha)^2 -{1\over \beta^2}
\big(1-{1\over \beta}\big)^2 \log(1-\beta)^2\Big\} \Bigg] \, .
\label{genmess}
\ee
Eqn.~\ref{generalf2} and Eqn.~\ref{generalf1} are the central results of
this work~\footnote{Our expression for $F_2$ in Eq. (\ref{generalf2}) can 
be further simplified--a result which will be discussed in a forthcoming 
paper~\cite{KKMV}.}. Since they are the most general possible expressions for 
arbitrary values of $Q^2$, $M^2$, and $\mu^2$, it is inevitable that they 
look complicated. We shall show in the following sub--section that 
they simplify considerably in the high $q^2$ limit.

\subsection{Structure functions in the limit $q^2\rightarrow \infty$.}
\vskip 0.15in

We shall now obtain the leading twist limits of Eq.~\ref{generalf1} 
and Eq.~\ref{generalf2}. In particular we will show that our structure 
function for $M^2\rightarrow 0$ and $q^2\rightarrow \infty$ is identical 
to the structure function obtained by integrating the light cone Fock 
distribution in Eq.~\ref{seafinal}. That this should be the case is a 
well known property of the leading twist structure functions~\cite{Jaffe,
EllisFurmPetr}. Further, we will recover the Callan--Gross 
result~\cite{CallGross} $F_1=F_2/2x$ in this limit.

Consider first our general formula for $F_2$ in Eq.~\ref{generalf2}. The
trace simplifies considerably when we put $M=0$. For $q_t\gg p_t,k_t$, 
we obtain
\be
M^{++}\longrightarrow 16 q_t^2 \left(p_t^2 + p_t\cdot k_t\right) 
\nonumber \, .
\ee
In the logarithm, the ratio $M_{p-q}^2 /M_{p+k-q}^2 \rightarrow 1$. Finally, 
in the denominator of the integrals, 
\be
\left(M_{p-q}^2 M_{p+k}^2 - M_{p+k-q}^2 M_{p}^2 \right) \longrightarrow
M_q^2 \left(M_{p+k}^2 - M_p^2\right) \nonumber \, .
\ee
Then putting these back into our general expression~\footnote{The 
contribution in the $p_t$ integral in the region $p_t\sim q_t$ is 
identical to that from $p_t<<q_t$. This provides a factor of 2 that 
must be taken into account.} we obtain
\be
F_2 = {\sigma N_c\over 2\pi^4} \int d^2 p_t \int {d^2 k_t\over 
{(2\pi)^2}} {\tilde \gamma} (k_t) \left[1-{(p_t^2 + k_t\cdot p_t)
\over (k_t^2+2k_t\cdot p_t)}\log\left({(k_t+p_t)^2\over p_t^2}\right)\right]
\, .
\ee
A comparison with Eq.~\ref{seafinal} immediately reveals that, setting
$k_t\leftrightarrow p_t$, and integrating the latter over $p_t$ upto
$q^2$ gives an identical result to the one above. Thus we have recovered
a well known, non--trivial leading twist result as the limit of our 
general expression for $q^2\rightarrow\infty$ and $M\rightarrow 0$.

Even though our general expression for the longitudinal structure 
function $F_1$ looked terribly complicated, in the limit considered here
it is remarkably simple. In this limit $\alpha,\beta \rightarrow 1$ and
hence Eq.~\ref{genmess} for ${\cal F} \rightarrow {1\over 3}$--a constant!
The product in the denominator of Eq.~\ref{generalf1},  
$M_{p-q}^2 M_{p+k-q}^2 \rightarrow q^4$ and cancels the $q^4$ factor outside
the integral. From the sum rule Eq.~\ref{rule}, we find remarkably that
the complicated integral vanishes and Eq.~\ref{generalf1} reduces to
\be
F_1 = {F_2\over 2x} \, .
\ee
The above is the well known Callan--Gross relation.

We should clarify the result obtained above to avoid confusion. The reader 
may note above that the deviation from the Callan--Gross relation vanishes 
as a power law as $q^2\rightarrow \infty$. On the other hand, it is well 
known in QCD~\cite{ZWT,BBDM} that the violations of the 
Callan--Gross relation only disappear logarithmically 
as $q^2\rightarrow \infty$. The apparent contradiction is resolved by 
one realizing that the logarithmic violations  at large $q^2$ in QCD come from 
diagrams where the sea quark emits a gluon (thereby violating Feynman's parton 
model helicity argument). These diagrams are of higher order in our picture 
and are therefore not included. In fact, the deviations from the
Callan--Gross relation of the sort discussed above (at small x) should die 
off faster than logarithmically at very large $q^2$ because for sufficiently 
large $q^2$, the violations of the Callan--Gross relation should come 
from precisely the diagrams not included here. At moderate $q^2$ however, the
contributions we have discussed above should be important.

\section{Summary}
\vskip 0.15in

In this paper we have used a classical theory of the gluon field to derive
expressions for Fock space distribution functions of quarks and structure
functions for deep inelastic scattering.  This theory is valid at small x
when the gluon density is large.  In this region, the coupling constant
evaluated at this density scale is small.  With this density scale denoted
by $\mu^2$, we have seen that when $q^2 >> \alpha^2 \mu^2$ and $q^2 >>
M^2$, where $M$ is the mass of the heavy quark being probed, all leading
twist results are reproduced.  For $q^2 \le \alpha^2 \mu^2$ or
$q^2 \le M^2$, we derived an expression valid to all 
orders in twist but only to leading order in $\alpha_S$.
In this kinematic region, the Callan-Gross relation is not
valid, and there is no simple relation between the Fock space distribution
function and the structure functions for quarks.

The structure function of heavy quarks deserves more study.  If $M^2
>> \alpha^2 \mu^2$, then for $q^2 >> \alpha^2 \mu^2$, the heavy quark
distribution is a linear function of the gluon distribution function.
The gluon distribution function may nevertheless be computed to all orders
in twist. The leading twist relation between the structure function and the
quark distribution function is however not valid for large $M^2$.
As we go to smaller values of $x$ corresponding to larger values
of $\mu^2$, the non-linearities corresponding to higher powers of the
gluon density turn on and can be studied
systematically in our weak coupling formalism.

The situation for light quarks is also amusing and needs more study as well.
If $q^2 \le \alpha^2 \mu^2$, then the non-linearities in the gluon
distribution function become important.  In this kinematic region,
we expect saturation of the gluon distribution function, and our function
$\gamma (k_t) \sim 1/k_t^2$ up to logarithms of $k_t$.  If this is the
case, a look at the definition of $F_2$ and $F_1$ shows that these
distributions should be dimensionally of order $q^2$.  In this region a
precise analytic estimate is difficult since in the integral
representations for $F_1$ and $F_2$ (Eqs.~\ref{generalf1} and
~\ref{generalf2} respectively), the there is no hierarchy of
momentum scales.  All momenta are of order $q$, and the integrand does not
simplify much.  Nevertheless, we see that the saturation of the gluon
density is sufficient to imply saturation of the quark density.

Both the study of the heavy quark and light quark distributions merit more
theoretical and phenomenological work within the framework described in
this paper.

\section*{Acknowledgements}
\vskip 0.1in
We dedicate this work to the memory of our friend and former colleague
Klaus Kinder--Geiger. We would both like to thank Leonid Frankfurt, 
Eugene Levin, and Mark Strikman for reading the manuscript and for their 
very useful comments and suggestions.
We would like to thank NORDITA (L.M) and TPI, 
Minnesota (R.V) for their hospitality. R.V.'s work was supported by the
Danish Research Council and the Niels Bohr Institute. L.M's work is 
supported by DOE grant DE-FG02-87ER40328.

\section*{Appendix A: Notation and Conventions}
\vskip 0.15in

We start by defining our convention and notations. Our metric is the
$+2$
metric $\hat{g}^{\mu\nu}= (-,+,+,+)$. The gamma matrices in space--time
co--ordinates are denoted by carets. In the chiral representation,
\begin{displaymath}
{\hat{\gamma}}^0=\left( \begin{array}{ccc}
0 & I\\
I & 0\\
\end{array} \right)\,\, ; \,\,
{\hat{\gamma}}^i=\left( \begin{array}{ccc}
0 & \sigma^i\\
-\sigma^i & 0\\
\end{array} \right)\,\, ; \,\,
{\hat{\gamma}}^5=\left( \begin{array}{ccc}
I & 0\\
0 & -I\\
\end{array} \right)\,\, ,
\end{displaymath}
and $\{\hat{\gamma}^\mu,\hat{\gamma}^\nu\} = -2\hat{g}^{\mu\nu}$.
Above, $\sigma^i, i=
1,2,3$ are the usual $2\times2$ Pauli matrices and $I$ is the $2\times
2$
identity matrix. In light cone co--ordinates, $\gamma^{\pm}
= ({\hat{\gamma}}^0\pm {\hat{\gamma}}^3)/\sqrt{2}$ and
$\{\gamma^\mu,\gamma^\nu\}
=-2g^{\mu\nu}$, where $g^{++}=g^{--}=0$,  $g^{+-}=g^{-+}=-1$ and
$g_{t_1,t_2}=1$
where $t_1,t_2=1,2$ here stand for the two transverse co--ordinates.
Note for
instance that in this convention $A_+ = -A^-$ and $A_t=+A^t$. Also, 
$q^2 = -2q^- q^+ + q_t^2$ hence a ``space--like'' $q^2$ implying large
space--like components would correspond to $q^2 >0$.

We now define the projection operators
\be
\alpha^\pm = {\hat{\gamma}^0\gamma^\pm \over \sqrt{2}} \equiv
{\gamma^\mp \gamma^\pm\over 2} \, ,
\ee
which project out the two component spinors $\psi_\pm = \alpha^\pm \psi$
~\cite{KogutSoper}.
Some relevant properties of the projection operators $\alpha^\pm$ are
\be
(\alpha^\pm)^2 = \alpha^\pm \,\, ; \,\, \alpha^\pm\alpha^\mp =0 \,\, ;
\,\, \alpha^+ +\alpha^- =1 \,\, ; \,\, (\alpha^\pm)^\dagger = \alpha^\pm
\,.
\ee
It follows from the above that $\psi_+ + \psi_- = \psi$. In the
following,
we will also use the familiar Dirac conventions $\beta =
{\hat{\gamma}}^0$
and $\alpha_\perp ={\hat{\gamma}}^0\gamma_\perp$.

The two component spinor is the dynamical spinor in the light cone 
QCD Hamiltonian $P_{QCD}^-$ and it is defined in terms of creation and 
annihilation operators as 
\be
\psi_+ = \int_{k^+>0} {d^3 k\over{2^{1/4} (2\pi)^3}} \sum_{s=\pm {1\over
2}} \left[ e^{ik\cdot x} w(s) b_s (k) + e^{-ik\cdot x}
w(-s) d_s^\dagger (k)
\right] \, .
\ee
Above $b_s (k)$ is a quark destruction operator and destroys a
quark with momentum $k$ while $d_s^\dagger (k)$ is an anti--quark
creation operator and creates an anti--quark with momentum k. Also above 
the unit spinors $w(s)$ are defined as 
\be
w({1\over 2}) =\left( \begin{array}{c}
0 \\
1\\
0\\
0\\ \end{array} \right)\,\, ; \,\,
w(-{1\over 2}) =\left( \begin{array}{c}
0 \\
0\\
1\\
0\\ \end{array} \right)\, ,
\ee
The creation and annihilation operators obey the 
equal light cone time ($x^+$) commutation relations 
\be
\{b_s (\vec{k},x^+), b_{s^\prime}^\dagger (\vec{k^\prime},x^+)\} =
\{d_s (\vec{k},x^+), d_{s^\prime}^\dagger (\vec{k^\prime},x^+)\} =
(2\pi)^3 \delta^{(3)} (\vec{k}-\vec{k^\prime}) \delta_{s s^\prime} \, .
\ee 
The above definitions ensure that the light cone QCD Hamiltonian can be
defined as $P_{QCD}^- = P_0^- + V_{QCD}$, where the non--interacting piece
of the Hamiltonian is defined as
\be
P_0^- = \int {d^3 k \over {(2\pi)^3}} \sum_{s = \pm {1\over 2}} 
{(k_t^2 + M^2)\over {2k^+}}\,\,\left(b_s^\dagger (k) b_s (k)+d_s^\dagger (k) 
d_s(k)\right) \, .
\ee
The definition of quark distribution functions is further discussed in the
text of section 3.2

The dynamical components of the gauge fields $A_i^a (x)$ with $i=1,2$ in 
light cone gauge $A^+=0$ are defined as
\be
A_i^a (x) = \int_{k^+>0} {d^3 k\over{\sqrt{2|k^+|}(2\pi)^3}} \sum_{\lambda =
1,2}\delta_{\lambda i}
\left[ e^{ik\cdot x} a_{\lambda}^a (k) + e^{-ik\cdot x}
{a_{\lambda}^a}^\dagger (k)\right] \, , 
\ee
where the $\lambda$'s here correspond to the two independent polarizations
and ${a_\lambda^a}\dagger (a_\lambda^a)$ 
creates (destroys) a gluon with momentum $k$. 
They obey the commutation relations
\be
[a_{\lambda}^a (\vec{k}), {a_{\lambda^\prime}^b}^\dagger 
(\vec{k^\prime})] =
(2\pi)^3 \delta^{(3)} (\vec{k}-\vec{k^\prime}) \delta_{ab} \delta_{\lambda 
\lambda^\prime} \, .
\ee
The gluon distribution function is then defined as
\be
{dN\over {d^3 k}} = {{a_i^a}^\dagger a_i^a\over {(2\pi)^3}} \, .
\ee
Performing the Fourier transform of the gauge field above, we obtain 
Eq.~\ref{correlator} in section 5.3
For a more extensive discussion of the above formalism 
see Ref.~\cite{capetown}.

We should mention here that there are several conventions in use.
For a discussion of
some of these, see the review article by Brodsky, Pauli and
Pinsky~\cite{BrodskyPauliPinsky}. Our convention is alike that of Kogut
and Soper~\cite{KogutSoper} but differs from theirs for quark 
spinor and gauge field normalizations by a factor 
$\sqrt{|k^+|}/\sqrt{(2\pi)^3}$.

\section*{Appendix B: Derivation of the Function ${\tilde \gamma}(p_t)$
for Gaussian Fluctuations}
\vskip 0.15in

Since ${\tilde \gamma}(p_t)$ is defined as the Fourier transform of
${1\over N_c}<{\rm Tr}(
U(x_t)U^\dagger (y_t))>_{\rho}$, we need to compute this correlator in
co--ordinate space first.  Note that the symbol $<\cdots >_{\rho}$
denotes
the averaging over with a Gaussian weight.
Now, as discussed by Jalilian--Marian et al., if
\be
U(y,x_t) = U_{\infty,y}(x_t) = {\rm P} \exp\left[ i \int_y^\infty
dy^\prime
\Lambda(y^\prime,x_t)\right] \, ,
\ee
where $U_{\infty,y}(x_t)$ is the path ordered exponential (in rapidity)
which
corresponds to the pure gauge potential $A^i=-U_{\infty,y}(x_t)
\nabla^i U_{\infty,y}^\dagger (x_t)/ig$, then $A^i$ satisfies the
Yang--Mills
equation
\be
D_i {dA^{i,a}\over {dy}} = g\rho^a (x_t,y) \, ,
\ee
if $\Lambda$, the argument of the path ordered exponential, satisfies
the
Laplace equation
\be
\nabla^2 \Lambda^a (x_t,y) = \rho^a (x_t,y) \, .
\ee

The measure for the functional integral is then
\be
\int [d\rho] \exp\left(-\int_0^\infty dy\int d^2 x_t {{\rm Tr}\rho^2
\over
\mu^2(y)}\right) \rightarrow \int [d\Lambda] \exp\left(-\int_0^\infty dy
\int d^2 x_t
{{\rm Tr} (\Lambda \nabla^4 \Lambda)\over {g^4 \mu^2(y)}}\right) \, .
\ee
As argued by Jalilian--Marian et al., we can write
\be
U_{\infty,y}(x_t) = :U_{\infty,y} (x_t): \exp\left(-{g^4 N_c
\Gamma(0)\over 2}
\int_y^\infty dy^\prime \mu^2 (y)\right) \, ,
\ee
where $:\cdots :$ denotes normal ordering and
\be
\Gamma (x_t) = {1\over \nabla^4}\equiv \Gamma(0) + {x_t^2\over
{16\pi}}\log(x_t^2 L^2)+finite\,\, pieces...  \, .
\label{bigam}
\ee
Above, $\Gamma(0)\propto 1/L^2$ where $L$ corresponds to an infrared
cut-off.
Writing $U_{\infty,y}(x_t)$ in the above normal ordered form enables us
to isolate
and exponentiate the infrared singular terms coming from disconnected
graphs.

Our correlator has then the form
\be
\gamma(x_t,{\bar x}_t; y, {\bar y}) =
N_c\gamma^\prime (x_t,{\bar x}_t; y, {\bar y})\exp\left(-{g^4 N_c
\Gamma(0)\over 2}\, \left(\int_y^\infty +\int_{{\bar y}}^\infty\right)\,
 dy^\prime \mu^2 (y^\prime)\right) \, ,
\ee
where
\be
\gamma^\prime (x_t,{\bar x}_t;y,{\bar y})=\int [d\Lambda] \exp\left(-
\int_0^\infty dy \int d^2 x_t
{{\rm Tr} (\Lambda \nabla^4 \Lambda)\over {g^4 \mu^2(y)}}\right)
\left(:U_{\infty,y} (x_t):\right) \left(:U_{\infty,y}
({\bar x}_t)\right)^\dagger \nonumber \, .\\
\ee
Expanding out first few terms in the path ordered exponentials above, we
have
\be
{\gamma^\prime}^{(0)} &=& 1 \, ,\nonumber \\
{\gamma^\prime}^{(1)} &=& {N_c^2-1\over {2 N_c}} \Gamma(x_t-{\bar x}_t)
g^4
\theta(y-{\bar y})\int_y^\infty dy^\prime \mu^2(y^\prime) \, , \nonumber
\\
{\gamma^\prime}^{(2)} &=& {1\over {2!}}{\left[{N_c^2-1\over {2N_c}}
\Gamma(x_t-{\bar x}_t) g^4 \int_y^\infty dy^\prime
\mu^2(y^\prime)\right]}^2
\, .
\ee
In the expression for ${\gamma^\prime}^{(2)}$ above only one of the two
possible terms survive on account of the path ordering. From similar
considerations it can be argued that in general
\be
{\gamma^\prime}^{(n)}={1\over {n!}}{\left[{N_c^2-1\over
{2N_c}}\Gamma(x_t-
{\bar x}_t) g^4\xi(y)\right]}^2 \, ,
\ee
where $\xi(y)=\int_y^\infty dy^\prime \mu^2(y^\prime)$. Resumming the
terms
above and including the disconnected pieces, we have
\be
\gamma(x_t-{\bar x}_t;y,{\bar y})=\exp\left({g^4 (N_c^2-1)\xi\over
{2N_c}}
(\Gamma(x_t-{\bar x}_t)-\Gamma(0))\right) \, .
\label{nonpertgam}
\ee
The above expression is the complete non--perturbative result for
$\gamma(x_t-{\bar x}_t;y,y^\prime)$.
Note that trivially $\gamma(0;y,{\bar y})=1$ as we would expect from the
definition of $\gamma$.

Taking the Fourier transform of $\gamma$,
\be
{\tilde \gamma} (p_t) &=& \int d^2 x_t e^{ip_t\cdot x_t}\gamma(x_t)
\nonumber \\
&\equiv& \int d^2 x_t e^{ip_t\cdot x_t} \exp\left[\kappa (\Gamma(x_t)-
\Gamma(0))\right] \, ,
\label{fougam}
\ee
where $\kappa = g^4 (N_c^2-1)\xi(y)/2 N_c$, and expanding out
$\gamma(x_t)$,
we obtain
\be
{\tilde \gamma}(p_t) = \left[\delta^{(2)}(p_t) + \kappa\int d^2 x_t
e^{ip_t\cdot
x_t} \Gamma(x_t) - \kappa \Gamma(0)\delta^{(2)}(p_t)+\cdots\right] \, .
\ee
If we now recall the definition of $\Gamma(x_t)$ from Eq.~\ref{bigam},
\be
\Gamma(x_t)=\int {d^2 k_t\over {(2\pi)^2}} {e^{ik_t\cdot x_t}\over
{k_t^4}}
\label{bigam2}
\, ,
\ee
and substitute for $\Gamma(x_t)$ in the above, we find
\be
{\tilde \gamma}(p_t) =
(1-\kappa\Gamma(0))\delta^{(2)}(p_t)+{N_c^2-1\over
{2N_c}} {(4\pi)^2\alpha_S^2\xi \over p_t^4}+\cdots \, .
\ee
The first term in ${\tilde \gamma}(p_t)$ is not relevant for our
computation
of the sea quark distributions since the relevant momenta are
$p_t>>\Lambda_{QCD}$. The
second term is the perturbative expression computed by us
previously~\cite
{MV2}--upto a factor of two which was missing in that paper. Note also
that the perturbative ${\tilde \gamma}(p_t)$ above explicitly satisfies
the
sum rule condition of Eq.~\ref{rule}.

\end{document}